# Pseudo-hydrodynamic flow of quasiparticles in semimetal WTe$_2$ at room temperature

Young-Gwan Choi[1]†, Manh-Ha Doan[1,2]†, Gyung-Min Choi[1,3]*, Maxim N. Chernodub[4,5]*

[1] Department of Energy Science, Sungkyunkwan University, Suwon 16419, Korea

[2] Department of Physics, Technical University of Denmark, Kgs. Lyngby 2800, Denmark

[3] Center for Integrated Nanostructure Physics, Institute for Basic Science, Suwon 16419, Korea

[4] Institut Denis Poisson UMR 7013, Université de Tours, 37200, France

[5] Pacific Quantum Center, Far Eastern Federal University, Sukhanova 8, Vladivostok, Russia

†These authors contribute equally
*Corresponding authors (gmchoi@skku.edu; maxim.chernodub@cnrs.fr)

**Recently, much interest has emerged in fluid-like electric charge transport in various solid-state systems. The hydrodynamic behavior of the electronic fluid reveals itself as a decrease of the electrical resistance with increasing temperature (the Gurzhi effect[1,2]) in narrow conducting channels[3,4], polynomial scaling of the resistance as a function of the channel width[5], substantial violation of the Wiedemann–Franz law[6,7] supported by the emergence of the Poiseuille flow[4,8-10]. Similarly to whirlpools in flowing water, the viscous electronic flow generates vortices, resulting in abnormal sign-changing electrical response driven by the backflow of electrical current[11]. Experimentally, the presence of the hydrodynamic vortices was observed in low-temperature graphene as a negative voltage drop near the current-injecting contacts[12]. However, the question of whether the long-ranged sign-changing electrical response can be produced by a mechanism other than hydrodynamics has not been addressed so far. Here we use polarization-sensitive laser microscopy to demonstrate the emergence of visually similar abnormal sign-alternating patterns in charge density in multilayer tungsten ditelluride at room temperature where this material does not exhibit true electronic hydrodynamics[13]. We argue that this pseudo-hydrodynamic behavior appears due to a subtle interplay between the diffusive transport of electrons and holes. In particular, the sign-alternating charge accumulation in WTe$_2$ is supported by the unexpected backflow of compressible neutral electron-hole current, which creates charge-neutral whirlpools in the bulk of this nearly compensated semimetal. We demonstrate that the exceptionally large spatial size of the charge domains is sustained by the long recombination time of electron-hole pairs.**

Tungsten ditelluride (WTe$_2$) is a remarkable representative of a family of non-magnetic layered transition-metal dichalcogenides, which provides a ubiquitous arena for investigation of diverse quantum effects including the quantum spin Hall effect[14-16] in a single-layered WTe$_2$ and a type-II Weyl semimetal phase in thick crystals[17,18].

The electronic transport in semimetallic WTe$_2$ at temperature 20 K demonstrates collective hydrodynamic behavior that appears to be in sharp contrast with the expected diffusive drift of a Fermi liquid[13]. The charge flow resembles a classical fluid when the momentum-conserving carrier-carrier scattering length becomes substantially shorter than the momentum-relaxing length governed by interactions of the charge carriers with phonons and impurities. Outside of a narrow temperature window (in particular, at room temperature), the charge transport in narrow WTe$_2$ flakes cannot be described as a hydrodynamic flow because the above condition is not fulfilled liquid[13].

The hydrodynamic flow of electrons can lead to a nonlocal macroscopic signature of viscosity[18-20]. In graphene, a localized injection of the electronic current creates whirlpools in the electron flow thus resembling a classical fluid[11]. The whirlpools drive the electric current against the applied electric field, resulting in a negative nonlocal voltage near narrow electrodes. The sign-flipping voltage pattern is an experimentally detectable signature of the electronic viscous behavior[12].

We experimentally observe visually similar sign-alternating patterns of charge accumulation close to the injection electrodes in a *film* of semimetallic WTe$_2$ with a thickness of about 70 nm *at room temperature* (see Methods and SM S1, S2). The observed structures in electronic charge density possess exceptionally long relaxation profiles of the micrometer length scale exceeding by three orders of magnitude the electrostatic screening length. Despite the bulk WTe$_2$ does not exhibit true electronic hydrodynamics at room temperature[13], we argue that the observed long-range patterns appear due to the formation of fluid-like whirlpools of a *neutral* quasiparticle current.

We induce charge accumulation on WTe$_2$ by applying an electrical bias. To visualize charge distribution, we use the optical microscopy technique (see Methods and SM S3). Charge accumulation affects the optical refractive index via the electro-optic effect[21,22]. A change of the refractive index in an anisotropic crystal leads to a rotation in the polarization ($\Delta\theta$) of the reflected light, which can be detected with a high precession of $10^{-7}$ radian (SM S4). To ascertain the sensitivity of the optical measurement of the accumulated charge, we induce a charge density in the capacitor structure of Si/SiO$_2$/ WTe$_2$ by applying a back-gate bias as shown in Fig. 1a. The optical penetration depth in WTe$_2$ is about 30 nm (ref. [23]) so that the probe beam can detect charge accumulation near the WTe$_2$/SiO$_2$ interface. A homogeneous $\Delta\theta$ is observed on the whole surface of the WTe$_2$ (Fig. 1b). The dependence of the polarization rotation $\Delta\theta$ on the incident angle $\theta$, shown in Fig. 1c, exhibits a sin $2\theta$ behavior without any offset thus showing that charge accumulation modulates the diagonal part of the dielectric tensor, $\Delta\varepsilon_{ii}$ (SM S5). Charge accumulation changes the Fermi level of WTe$_2$, which, in turn, modifies the interband transition rate, and, consequently, modulates the diagonal part of the dielectric tensor, $\Delta\varepsilon_{ii}$ (SM S5). Thereby, we establish a linear relationship between the $\Delta\theta$ and the charge accumulation, $\Delta\theta \propto \Delta\varepsilon_{ii} \propto \Delta n$, where $n$ is the charge density[21,22].

An entirely different distribution of charge emerges when electrical current is injected at the edges of the WTe$_2$ device through two long metal electrodes (Fig. 1d). We clearly observe the $\Delta\theta$ signals near the longitudinal boundaries of the devices, i.e., near the Au/WTe$_2$ interface with an opposite sign at an opposite boundary (Fig. 1e). It is worth noticing that the charge accumulation possesses exceptionally long spatial profiles with a characteristic length of $\lambda_a = 1.4$ $\mu m$ along the $a$-axis (Fig. 1f and SM S6). The magnitude of $\Delta\theta$ is linearly proportional to current density ($J$) with the same $\lambda_a$ (inset of Fig. 1f). A similar measurement along the $b$-axis produces a slightly different value $\lambda_b = 0.7$ $\mu m$ highlighting the anisotropic nature of WTe$_2$ (SM S7). These lengths are anomalously long given the fact that the WTe$_2$ semimetal has a high carrier concentration, $\sim 10^{20}$ cm$^{-3}$ (ref. [24]), which leads to a short screening length smaller than 1 nm (ref. [25]). We argue that the charge accumulation is driven by the flow of quasiparticles (both electrons and holes). The size of the charge accumulation region is determined by the recombination length, $\sim 1$ $\mu$m, of the electrons and holes. We argue below that the quasiparticle flow is responsible for both the long-range charge accumulation and the pseudo-hydrodynamic behavior.

To clarify the origin of the long-relaxing charge distribution, we fabricate a device that has four narrow electrodes a width of 2.5 $\mu$m attached to the edges of a WTe$_2$ flake. The electrodes are aligned along $a$- and $b$-axis as shown in Fig. 2a and d, respectively. Surprisingly, our optical measurements indicate the emergence of unexpected negative-positive-negative patterns of charge accumulation in the proximity of the electrodes. A visually similar nonlocal sign-changing voltage pattern reflects the hydrodynamic behavior of electrons caused by a backflow of emergent electronic whirlpools in graphene[11]. In this material, the hydrodynamic response was experimentally observed as a negative voltage drop in the vicinity of the contacts[12]. We argue that the observed sign-alternating charge pattern in the room-temperature

WTe$_2$ is caused by similar whirlpools of an *electrically neutral* quasiparticle current of electrons and holes, associated with a pseudo-hydrodynamic behavior in this two-carrier material.

WTe$_2$ is a nearly compensated semimetal that hosts both types of charge carriers, electrons, and holes. In the disorder-dominated regime, the two-carrier models admit simple kinetic descriptions with deep consequences for transport properties in confined geometries[26-28]. The electronic ($\alpha = e$) and hole ($\alpha = h$) currents $\boldsymbol{j}_\alpha$, and the corresponding charge densities, $n_\alpha(r) = n_{0,\alpha} + \delta n_\alpha(r)$, obey, in the background of the electric field $\mathbf{E} = -\boldsymbol{\nabla}\phi$, the steady-state transport relations (SI 8),

$$D_\alpha \boldsymbol{\nabla} \delta n_\alpha - \text{sign}(e_\alpha)\sigma_\alpha \mathbf{E} = -\boldsymbol{j}_\alpha, \qquad (1)$$

where $D_\alpha$ and $\sigma_\alpha$ are the diffusion constants and the normalized conductivities, respectively. The electron and hole currents are not separately conserved due to the e-h recombination process characterized by relatively long recombination time $\tau_R$.

The charge density $\delta n = \delta n_e - \delta n_h$ satisfies the following fourth-order differential equation (SI 9):

$$(\Delta - \lambda_1^{-2})(\Delta - \lambda_2^{-2})\delta n(x) = 0, \qquad (2)$$

where $\Delta = \partial_x^2 + \partial_y^2$ is the two-dimensional Laplacian appropriate for a thin-film geometry. In a near-neutrality regime, the length parameters in Eq. (2) are related, respectively, to the short Thomas-Fermi length $\lambda_1 \approx \lambda_{\text{TF}} \approx 1$ nm and much longer electron-hole recombination length $\lambda_2 \approx \lambda_{\text{R}} \approx 1$ $\mu m$.

One notices that charge-density structures characterized by a length scale shorter than the recombination length $\lambda_R$, Eq. (2) acquires a formal mathematical resemblance with the hydrodynamic relation[11],

$$\Delta \left(\Delta - \frac{1}{D_\nu^2}\right)\psi = 0, \qquad (3)$$

imposed on the stream function $\psi$ in a real electronic fluid system. The stream function describes the electronic velocity $\mathbf{u} = \hat{\mathbf{z}} \times \boldsymbol{\nabla}\psi$ of a two-dimensional incompressible electric current $\mathbf{J} = -e\mathbf{u}$ in a one-component conductor (here $\hat{\mathbf{z}}$ is a unit out-of-plane vector). Equation (3) can be derived from the linearized Navier-Stokes equation which describes, for example, the non-Ohmic electron flow $\mathbf{J} = -e\mathbf{u}$ in graphene[11]:

$$\left(\frac{\sigma_0}{e}\right)\boldsymbol{\nabla}\phi(r) - D_\nu^2 \Delta\mathbf{u}(r) + \mathbf{u}(r) = 0, \qquad (4)$$

with conductivity $\sigma_0$. The quantity $D_\nu \propto \sqrt{\eta}$ plays a role of a diffusion constant expressed via the electronic shear viscosity $\eta$. At zero viscosity, $\eta = 0$, the diffusion constant vanishes, $D_\nu =$ 0, and the Navier-Stokes equation (4) reduces to the usual Ohmic flow $\mathbf{J} = \sigma_0 \mathbf{E}$. The non-harmonicity of the charge accumulation (2) in the two-component diffusive system (1) resembles non-harmonic behavior (3) of the viscous non-Ohmic electric flow (4) in the one-carrier viscous electronic system such as graphene[11-12].

An unexpected analogy of the diffusive two-component system (1) with hydrodynamics can also be established in the opposite limit of large distances where the geometric features of charge accumulation domains are set by the recombination length $\lambda_R$. In this regime, the pseudoparticle current, $\mathbf{u} \equiv \mathbf{P} = \mathbf{j}_e + \mathbf{j}_h$, is described by the suggestive linearized Navier-Stokes equation (SM S9):

$$-\bar{\eta}\Delta\mathbf{u} - \left(\bar{\zeta} + \frac{\bar{\eta}}{3}\right)\nabla(\nabla\cdot\mathbf{u}) = \bar{\mathbf{f}} - \frac{1}{\tau_{\mathrm{mr}}}\mathbf{u} + \cdots . \qquad (5)$$

Here the ellipsis denotes the higher-derivative (Burnett-like[29]) terms that are irrelevant in the long-distance limit, $l \gg \lambda_R$, but important to maintain the potential nature of the *non-conserved*, compressible flow $\mathbf{P}$. If the quasiparticle current $\mathbf{P}$ were conserved, then the lowest-derivate terms in (5) would acquire the hydrodynamic form (4). As we show below, at these large distances, the system (1) supports the formation of whirlpools and backflow of the neutral quasiparticle current, hence we call this regime "pseudo-hydrodynamics".

In Eq.(5), the kinematic shear viscosity $\bar{\eta} \equiv \lambda_+^2/\tau_R$ is expressed via the length parameter $\lambda_+ = \sqrt{(D_e + D_e)/(8\pi e(\sigma_e + \sigma_h))}$ which also enters the kinematic bulk viscosity $\bar{\zeta} \equiv (D_e + D_e)/2 - \lambda_+^2/(3\tau_R)$. The electron-hole recombination time $\tau_R$ is identified with the momentum-relaxation time $\tau_{\mathrm{mr}} \equiv \tau_R$ related to disorder scattering while the force acting on the neutral quasiparticles is $\bar{\mathbf{f}} \equiv \frac{\sigma_e - \sigma_h}{\tau_R}\mathbf{E}$.

The whirlpools in the electrically neutral quasiparticle current $\mathbf{P}$ cannot be detected experimentally in a direct way. However, outside of the neutrality point, the quasiparticle flow is known to lead[27,28] to accumulation of local electric charge density $\delta n(\mathbf{r})$ at the boundaries (SM S10) which can detectable experimentally. Our optical data allows us to reconstruct the quasiparticle current via theoretical analysis.

In Figs. 2c,f, we show the theoretical results for the charge distribution corresponding to the experimentally realized geometry of Figs. 2a and d, respectively, with the two pairs of electrodes attached at the opposite edges of the sample. We obtain the charge density accumulation by solving the continuity and Maxwell equations supplemented by the transport relations (1), with appropriate boundary conditions (the method is described in detail in Supplementary Materials S12). We took the width of the electrodes $b = 2.5\ \mu m$ and set the distance between the electrodes $L = 15 \ \mu$m. The difference in the recombination lengths in the $a$ and $b$ axes of the crystal has no qualitative effect on the charge accumulation except for a global re-scaling.

The almost symmetric charge accumulation pattern (Fig. 2a,b), generated by the current injection to the vertical electrodes, with the zoom-in of the experimental data given in Fig. 2b, agrees very well with the prediction coming from the theoretical model with symmetrically positioned electrodes (Fig. 2c). The theoretical model excellently describes all seven regions of the charge-alternating pattern including three sign-changing regions close to each of the electrodes and the vanishing charge in the central region of the sample.

The asymmetric charge accumulation pattern (Fig. 2d,e), generated by the current injection to the horizontal electrodes, can be reproduced theoretically by introducing a vertical mismatch in 3 $\mu$m between the electrode centers (Fig. 2f). The two-carrier theory shows that the experimentally observed picture arises at any non-negligible shift between the electrodes. Contrary to the seven disentangled regions produced in the symmetric arrangement of the electrodes, Fig. 2a, the mismatch reduces the number of the same-sign charge regions to four, Fig. 2d. The upper and lower accumulation regions are separated by the wiggling boundary curve which stretches in the predominantly horizontal direction between the electrodes. In the experiment, this curve has a smaller amplitude compared to the theoretical prediction, presumably due to imperfections in the electrode positions. The electric charge accumulation shown in Figs. 2a and 2d is the experimentally detectable part of a more complicated picture emerging in imbalanced two-carrier (semi)metals: the neutral particle current necessarily leads to a neutral charge accumulation at the boundaries which generates, in turn, the large domains of nonvanishing electric charge density (SM 10-13).

The neutral quasiparticle flow **P** contains, besides a featureless Ohmic part, a nontrivial non-Ohmic part that leads to the experimentally observable electric charge accumulation. The neutral flow, shown in Fig. 3, possesses several remarkable (pseudo-)hydrodynamic features which are usually attributable to a collective fluid-like behavior rather than to the disorder-dominated transport. Unexpected in WTe$_2$ at the room-temperature, the hydrodynamic-like behavior manifests itself via the backflow of the neutral current at the corners of both electrodes which leads to the appearance of the sign-alternating regions of the accumulated electric charge of Fig. 2c,f. The picture is strikingly similar to the viscous behavior of the electronic fluid in graphene, where the sign-alternating electrostatic potential near the contacts is generated by the fluid backflow. In WTe$_2$, the sign-alternating charge density is generated by the compressible backflow of non-conserved neutral quasiparticle currents which terminate at the boundaries of the sample and generate, outside of the exact particle-hole compensation regime, the nonvanishing electric charge density. It is worth stressing that the quantities which exhibit the charge-alternating patterns, the underlying mechanisms that generate these patterns, and the temperature regimes are different in graphene and $WTe_2$.

The pseudo-hydrodynamic picture of neutral currents is supported by two well-pronounced symmetric whirlpools of the neutral quasiparticle flow, Fig. 3a. The whirlpool cores are localized in the regions with depleted electric charge density. The misaligned electrodes also generate the whirlpools, which possess sink/source attractor-like points in their cores, Fig. 3b. These whirlpool structures share similarities with electronic vortices in graphene: they correlate with the quasiparticle backflow which generates the sign-alternating charge density near the contacts, Fig. 3c.

The whirlpool size is determined by the recombination length $\lambda_R \sim 1$ $\mu$m which is typical for (nearly-)compensated semimetals[27]. The very existence of the whirlpools is largely insensitive to the value of the electrostatic (Thomas-Fermi) screening length $\lambda_{TF}$ provided $\lambda_{TF} \ll \lambda_R$. Both our experimental data and the results of the theoretical model demonstrate that the sign-alternating charge patterns (Fig. 2) and the positions of the neutral whirlpools (Fig. 3) and can be controlled by the sizes and positions of the electrodes.

This work opens the possibility of observing hydrodynamic-like effects of neutral quasiparticle currents in a wide range of two-component nearly compensated materials in which the true hydrodynamics regime of charge carriers cannot be realized. The comparison of theoretical predictions and experimental observations of accumulation patterns via polarization-sensitive laser microscopy allows us to uncover, at room temperature, the existence of the steady-state whirlpools and the backflow of the neutral currents in $WTe_2$. Our approach allows us to explore and quantify the pseudo-hydrodynamic neutral flows, suggesting an intriguing possibility to control and manipulate the charge accumulation patterns in (semi)metallic materials, which could be useful in future microelectronics.

# Methods

*Device fabrication:* WTe$_2$ multilayer with a thickness about 70 nm is exfoliated onto SiO$_2$/Si substrate by using the scotch tape method inside a glove box with an argon environment to prevent oxidation. The sample is then coated by a PMMA layer before being taken out for device fabrication. The thickness and the in-plane crystal orientation (*a*- and *b*-axes) of the exfoliated WTe$_2$ flake is determined by using atomic force microscopy and an angle-dependent polarized Raman spectroscopy, respectively (SM S1). The Cr/Au metal electrodes are fabricated along these two axes by the e-beam lithography and e-beam evaporation processes. The device is then etched by a reactive ion etching system using SF$_6$ gas to form a well-defined shape and coated by a thin SiN passivation layer to prevent surface degradation of WTe$_2$ during the electrical and optical measurements. The temperature dependence and crystal-axis dependence of the electrical resistivity manifest the metallic behavior of WTe$_2$ (SM S2).

*Optical measurement of charge accumulation:* A linearly polarized laser beam is incident onto WTe$_2$ devices in a surface normal direction with a center wavelength of 780 nm. After passing through a 50x objective, the beam radius is about 1.4 $\mu$m, and the input power density is about 1 mW/$\mu$m$^2$. The charge accumulation on WTe$_2$ rotates the polarization of the reflected light. The polarization angle of the incident laser beam was controlled by using a motorized rotational stage and a half-wave plate. The reflected laser beam is passing through a quarter-wave plate, half-wave plate, and Wollaston prism successively, and its polarization rotation is detected by a balanced detector. With the initial polarization along the azimuthal angle of $\theta$ to the *a*-axis of WTe$_2$ (the inset in Fig. 1c), we measure the current-induced polarization rotation of the reflected light, $\Delta\theta = \theta(J) - \theta(J = 0)$ (SM S3). To enhance the signal-to-noise ratio, a lock-in detection technique is implemented by applying an alternating current to WTe$_2$ devices at a modulation frequency of 3 kHz. The 2D spatial mapping of the optical signals is conducted by using a 2-axes motorized stage.

*Methods for theoretical calculations are described in Supplementary Information (Sections 8-13).*

**Acknowledgments**

Y.-G.C. and G.-M.C. are supported by Samsung Research Funding & Incubation Center of Samsung Electronics under Project Number SRFC-MA2001-03. This work was supported in part by Advanced Facility Center for Quantum Technology. The work of M.C. was partially supported by Grant No. 0657-2020-0015 of the Ministry of Science and Higher Education of Russia. G.-M.C. thanks Jungcheol Kim and Hyeonsik Cheong at Sogang University for their help in Raman Spectroscopy.

**Author contributions**

Y.-G.C. and M.-H.D equally contributed to this work. M.C. and G.-M.C supervised the study. Y.-G.C. performed the optical measurement and analysis. M.-H.D and Y.-G.C. fabricated samples and characterized optical and electrical properties. M.C. developed the theory part of the paper and participated in the writing of the text.

**Competing interests**

The authors declare no competing interests

**Additional Information**

Supporting information is available for this paper.

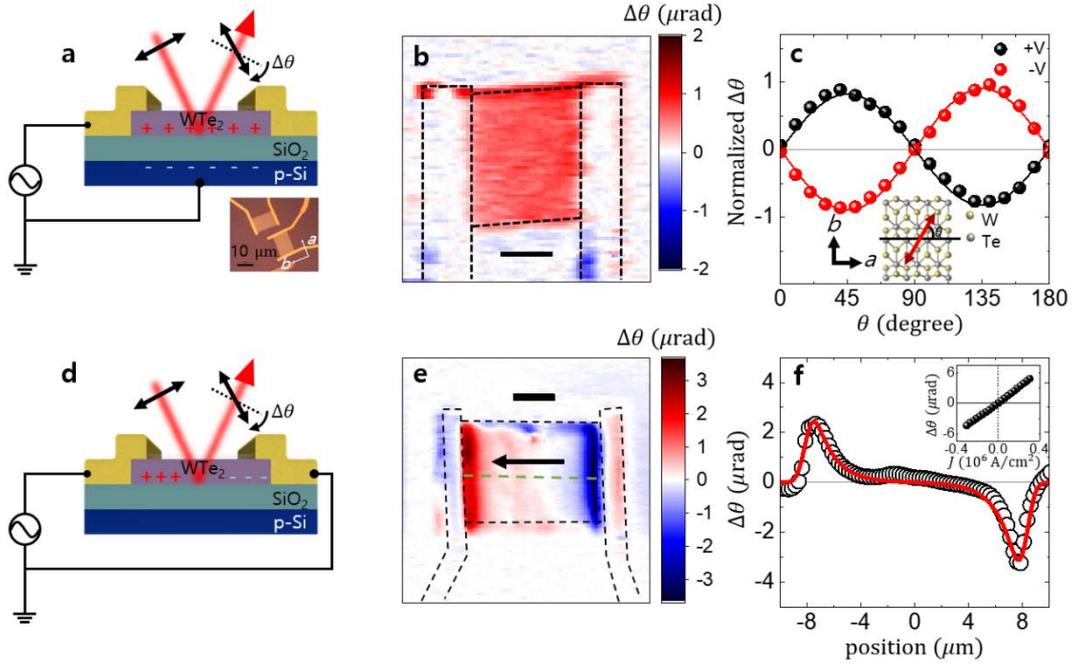

**Fig. 1 | Optical detection of long-range charge accumulation.** **a**, Measurement setup and **b**, the spatial mapping of the polarization rotation ($\Delta\theta$) with a capacitor structure of p-Si/SiO$_2$/WTe$_2$. An electric bias of 3 V induces charge accumulation at the WTe$_2$/SiO$_2$ interface. The charge accumulation rotates the polarization of the probe light (red arrow). The red/blue color in (b) shows the sign and the magnitude of the polarization rotation angle $\Delta\theta$. **c**, A sinusoidal dependence of $\Delta\theta$ on the azimuthal angle ($\theta$) between the light polarization and $a$-axis of WTe$_2$. The inset shows a top-view ($a$-$b$ plane) of the crystal structure of WTe$_2$. **d**, Measurement setup and **e**, the spatial mapping of $\Delta\theta$ with a lateral structure of Au/WTe$_2$/Au. An electric current of 2 mA flows along the $a$-axis of WTe$_2$, and the charge accumulates at the Au/WTe$_2$ boundary. The black bar in (b) and (e) indicates the length scale 5 $\mu$m. The black arrows point to the direction of the current flow. **f**, The profile of $\Delta\theta$ along the channel position, green dashed line in (e). The inset of (f) is $\Delta\theta$ at the channel edge as a function of current density. The $\Delta\theta$ is linear with the current density, and it is reversed when the charge current is reversed.

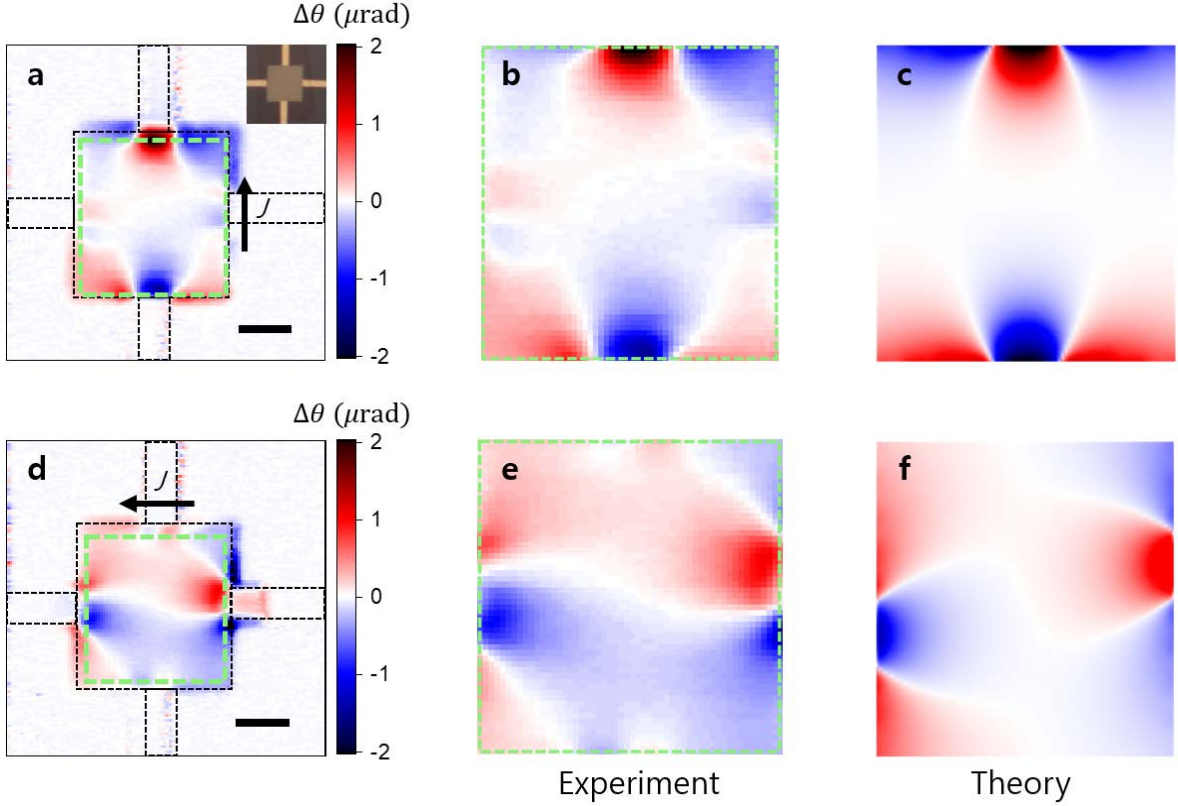

Experiment Theory

**Fig. 2 | Sign-alternating charge accumulation in 2D devices with narrow electrodes. a**, The spatial mapping of the polarization rotation ($\Delta\theta$) with the electric current of 2 mA along the *a*-axis of WTe$_2$. The red and blue colors indicate the sign and magnitude of $\Delta\theta$. **b**, Enlarged mapping image of green dashed line in (a). **c**, Theoretical simulation of the charge accumulation, which corresponds to experimental data of (a). **d,e**, The spatial mapping of $\Delta\theta$ and **f**, theoretical simulation of the charge accumulation with the electrical current along the *b*-axis. The black horizontal bars in (a) and (d) indicate the length scale 5 $\mu$m. The theoretical simulations match closely the experimental data in a wide range of recombination lengths $\lambda_R \simeq (0.5 - 2.0)$ $\mu$m depending on the boundary conditions (SM S13).

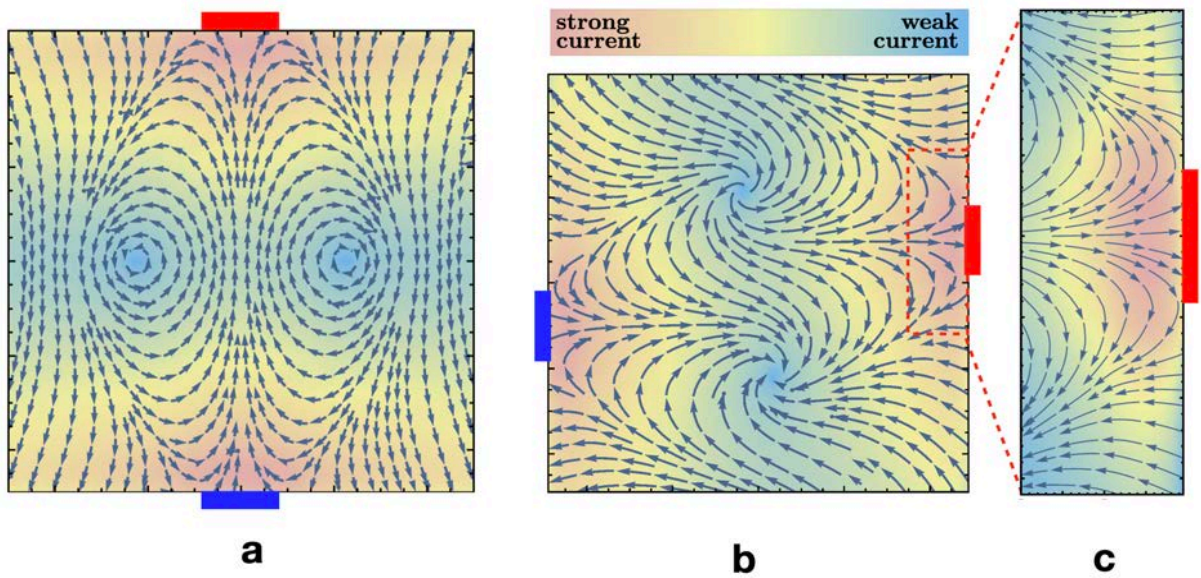

**Fig. 3 | Streamlines of the non-Ohmic component of the neutral quasiparticle flow P.**
**a**, The streamlines of theoretically calculated neutral current generated by the voltage difference between perfectly aligned electrodes. This figure corresponds to the accumulated electric charge density of Fig. 2c which matches, in turn, the experimental data of Fig. 2a. The color represents the magnitude of the quasiparticle current (shown in arbitrary units). **b**, Same as in (a) but for mismatched electrodes, corresponding to the charge accumulation of Fig. 2f which matches, in turn, the experimental data of Fig. 2d. **c**, the zoom-in region close to the rightmost contact (with colors enhanced to reveal more details). The positions of the electrodes are denoted by the solid rectangles.

# Supplementary Material

Pseudo-hydrodynamic flow of quasiparticles in semimetal WTe$_2$ at room temperature

Young-Gwan Choi[1]†, Manh-Ha Doan[1,2]†, Gyung-Min Choi[1,3]*, M. N. Chernodub[4,5]*

[1] Department of Energy Science, Sungkyunkwan University, Suwon 16419, Korea

[2] Department of Physics, Technical University of Denmark, Kgs. Lyngby 2800, Denmark

[3] Center for Integrated Nanostructure Physics, Institute for Basic Science, Suwon 16419, Korea

[4] Institut Denis Poisson UMR 7013, Université de Tours, 37200, France

[5] Pacific Quantum Center, Far Eastern Federal University, Sukhanova 8, Vladivostok, Russia

†These authors contribute equally

*Corresponding authors

# CONTENTS

# Supplementary Material (Experiment)

## S1. Characterization of the thickness and crystal orientation of $WTe_2$

To measure the thickness of the $WTe_2$ flakes, the atomic force microscopy (AFM) technique is used. It shows that the thickness of the $WTe_2$ device is about 70 nm (Fig. S1a,b). We observe that the *a*-axis is always oriented along the well-defined edge as shown in Fig. S1c, which is naturally formed after exfoliation processes as reported previously[1, 2]. To exactly determine the crystalline axis, polarization-dependent Raman spectroscopy is used. The polar Raman measurements were conducted with a confocal micro-Raman system at room temperature. The excitation source is the 1.96 eV (632.8 nm) line of a He-Ne laser. The laser beam was focused by a 50x objective lens (0.8 NA) onto the sample flake and the scattered light was collected (a backscattering geometry). Fig. S1d,e shows the polarization angle-dependent Raman results for our thin film of $WTe_2$ exfoliated on $SiO_2$/Si substrates. The Raman peak around 162 cm$^{-1}$ has a maximum intensity when the light polarization direction is aligned with the *a*-axis[1,2].

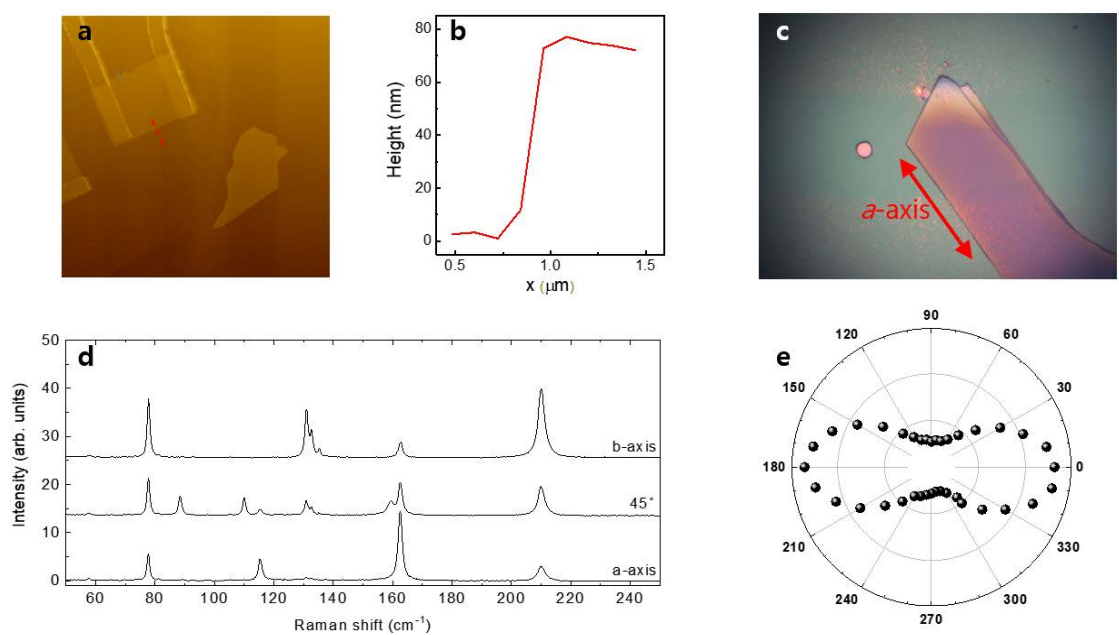

**Figure S1. a**, Optical microscopy image of the $WTe_2$ device. **b**, Atomic force microscopy result of fabricated devices and height line profile along the red dashed line in (a). **c**, Optical microscopy image of the $WTe_2$ flake before making devices. **d,** Raman spectra with three different light polarization angles, along the *a*-axis, 45° to *a*-axis, and along the *b*-axis. **e**, Polarization dependence on the Raman intensity at the 162 cm$^{-1}$ peak. It shows that the maximum intensity occurs when the light polarization is oriented along the *a*-axis.

## S2. Electric transport properties of WTe₂

The electrical characterization of the fabricated device is conducted at room temperature in a high vacuum (< 10⁻⁶ Torr) using a probe station with a semiconductor analyzer (Keithley 4200 system). Temperature-dependent transport measurements are carried out in a Physical Property Measurement System (PPMS-Quantum Design). The linear current-voltage characteristic and the decrease of resistance with a temperature of the fabricated devices manifest the metallic behavior of WTe₂ thin film[3-5]. The resistance of the *b*-axis device is larger than that of the *a*-axis device about 1.6~1.8 times at room temperature, which reflects the highly anisotropic property of WTe₂, and is in line with previous reports[6,7].

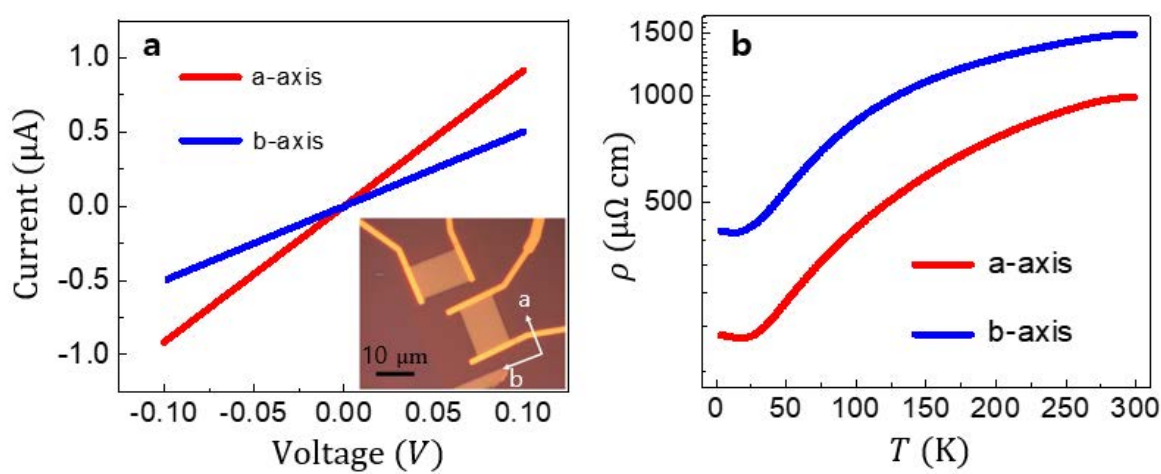

**Figure S2 a,** Current-voltage (IV) characteristics and **b**, temperature-dependent resistivity of WTe₂ with a current along the *a*- and *b*-axes of devices. The linear behavior of the IV characteristic is a manifestation of the Ohmic contacts. The decrease of the resistivity with decreasing temperature shows the metallic property of WTe₂. The inset in (a) shows an optical image of the *a*- and *b*-axes of devices.

## S3. Optical setup for measuring the polarization rotation

An optical setup in Fig. S3 was used for detecting polarization rotation of the probe light reflected from the sample surface with alternating current (AC) injection into the sample. As a light source, Ti:Sapphire pulsed laser was used with a 780 nm center wavelength. A current source is used for the application of AC to WTe₂ devices with a 3 kHz modulation frequency. The laser beam was incident onto the WTe₂ devices by passing through a 50x objective lens at the surface normal direction. The beam waist is about 1 *μ*m, and power is about 5 mW. For adjusting the azimuth angle of the light polarization, a half-wave plate (HWP1) was used. After reflection from the sample, the reflected light is separated from the incident light by a beam splitter (BS), passes through a half-wave plate (HWP2), then goes to a balanced detector. Given the complex nature of the dielectric constant of WTe$_2$, we measured the complex polarization rotation angle ($\Delta\tilde{\theta}$) with and without the quarter-wave plate (QWP), which is in between BS and HWP2, to measure the real part Re[$\Delta\tilde{\theta}$] and imaginary part Im[$\Delta\tilde{\theta}$], respectively. For all mapping measurements in the main text, Im[$\Delta\tilde{\theta}$] is shown because it is much larger than Re[$\Delta\tilde{\theta}$]. In the main text, $\Delta\theta$ denotes Im[$\Delta\theta$] for simplicity. A lock-in amplifier is used for collecting a subtle change of the polarization rotation of reflected light with the same modulation frequency of the AC.

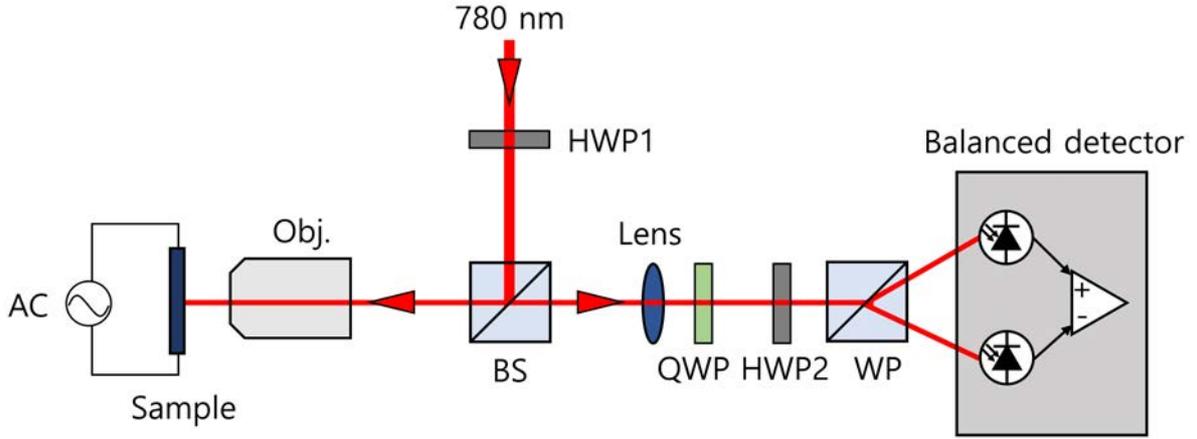

**Figure S3.** Schematic illustration of the AC-induced polarization rotation measurement. Obj. is the objective lens, BS is 5:5 beam splitter, HWP is half-wave retarder plate, QWP is quarter-wave retarder plate, and WP is Wollaston prism.

## S4. Noise level of the optical measurement

In our optical system, the dominant noise source is the intensity fluctuation of our laser. We measure the relative intensity noise (RIN) from the laser fluctuation as,

$$\text{RIN} = \frac{\text{spectral } V_{\text{rms}}}{V_{\text{DC}}}, \tag{S1}$$

where the spectral $V_{\text{rms}}$ is the voltage reading on the lock-in amplifier by the laser fluctuation per $\sqrt{\text{Hz}}$, and $V_{\text{DC}}$ is the DC voltage by the average laser power. The repetition of pulsed laser is injected into a normal Si photodetector. The photocurrent output from the photodetector is connected to the input of the lock-in amplifier. The $V_{\text{rms}}$ is measured at reference frequencies of 10~100,000 Hz. The $V_{\text{DC}}$ is determined as $V_{\text{DC}} = I_{\text{light}} \times R_{\text{detector}} \times Z_{\text{lock-in}}$, where $I_{\text{light}}$ is the average power of the laser, $R_{\text{detector}}$ is the responsivity of the photodetector of 0.4 A W$^{-1}$, and $Z_{\text{lock-in}}$ is the input impedance of the lock-in amplifier of 50 Ω. The measured RIN decreases with frequency (Fig. S4(a)); then, at >kHz frequency, saturates to a few 10$^{-6}$. When we measure the Kerr rotation using a balanced photodetector (Thorlabs, PDA450A-AC), the noise level further decreases by the common-mode rejection of the balanced detector. The typical noise level of our measurement at a bandwidth of 0.05 Hz is $5\times10^{-8}$, shown in Fig. S4b, determined from the standard deviation of multiple measurements. Further reduction of the noise level can be achieved by averaging multiple measurements.

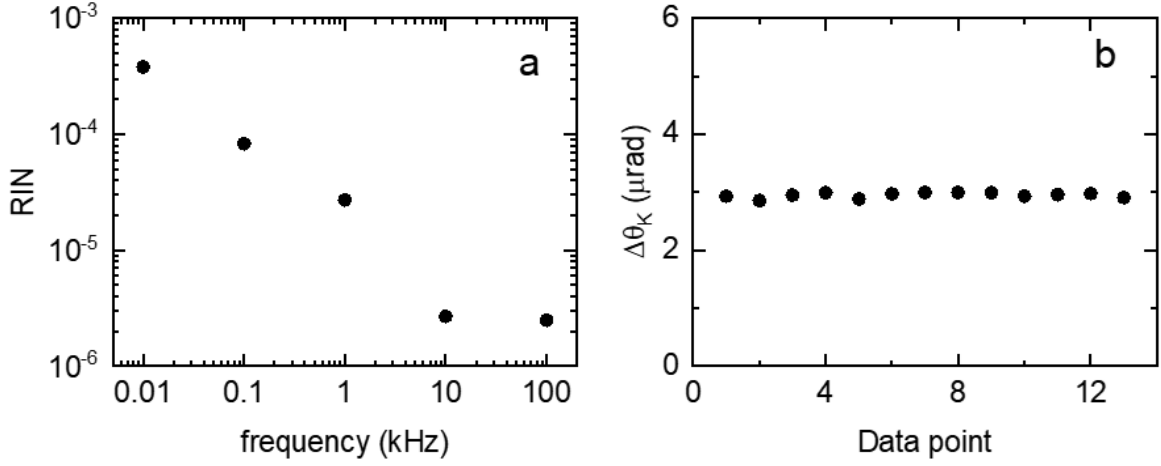

**Figure S4. a**, The relative intensity noise of the pulsed laser at different frequencies. **b**, Repeated measurements of the Kerr rotation at the same experimental condition. One data point is taken at the bandwidth of 0.05 Hz. The standard deviation of 13 data points is 0.05 µrad.

## S5. Modulation of the diagonal part of the dielectric tensor

WTe$_2$ is an optically anisotropic material with a reflectivity anisotropy of $\left|\frac{r_{yy}}{r_{xx}}\right|^2 \approx 1.2$, where $r_{xx}$ and $r_{yy}$ are the reflection coefficients for the light's polarization along the *a*- and *b*-axes, respectively (Fig. S5a). When the charge accumulation modulates the Fermi level, a change in the optical transition rate leads to a change in the dielectric constant of WTe$_2$ (Fig. S5b). Then, the modulation of the dielectric constant in an optically anisotropic medium results in a polarization rotation upon the reflection[8].

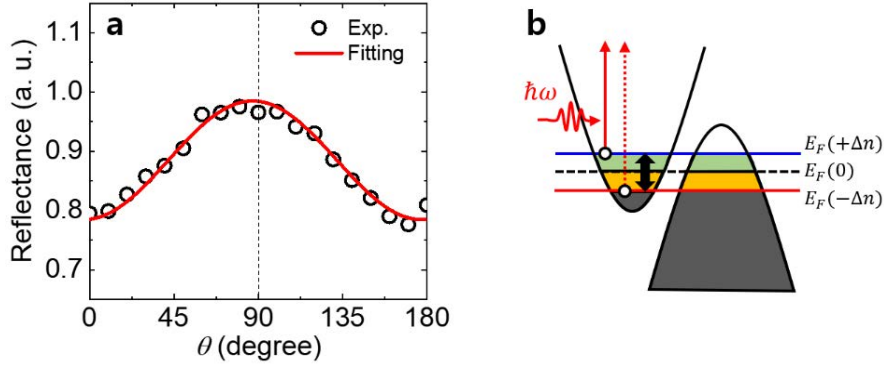

**Figure S5. a,** Azimuthal angle dependence of the reflectance, which is measured in the absence of the electric current. It shows anisotropic reflectance behavior. **θ** denotes an angle between light polarization direction and *a*-axis of the WTe₂ crystal. **b,** A schematic diagram for the change of Fermi level and interband transition owing to the charge accumulation. The red solid (dashed) arrow indicates optical interband transition when positive (negative) bias is applied.

The optical signal can be generated either by the diagonal or off-diagonal part of the dielectric tensor. To clarify which part is responsible for the charge accumulation effect, we calculate the polarization rotation by adopting the Jones matrix formalism. The polarization of light at the detector can be written as,

$$\begin{pmatrix} E_{\text{out}}^x(j_c) \\ E_{\text{out}}^y(j_c) \end{pmatrix} = M_{\lambda/2}\left(22.5° + \frac{\theta}{2}\right) M_{\lambda/4}(\theta)\, R(j)\, M_{\lambda/2}\left(\frac{\theta}{2}\right) \begin{pmatrix} E_{\text{in}}^x \\ E_{\text{in}}^y \end{pmatrix}, \qquad (S2)$$

where $\theta$ is the fast axis angle of the wave plates, $E_{\text{in}}^i$ ($E_{\text{out}}^i$) is the initial (final) electric field of light with polarization along $i$-direction, $j$ is electric current density, $M_{\lambda/2}$ ($M_{\lambda/4}$) are the matrix for the half-(quarter-)wave plate, and $R$ is the reflection matrix of WTe₂. The polarization of the initial light and the *a*-axis of WTe₂ are set to the *x*-direction. After passing through the first half-wave plate of $M_{\lambda/2}\left(\frac{\theta}{2}\right)$, the polarization of light (right before the sample) rotates by the angle $\theta$ with respect to the *x*-direction. After reflecting from WTe₂, the polarization of light rotates to $\theta+\Delta\theta$, where $\Delta\theta$ is polarization angle variation which is linearly proportional to $j$ (Fig. 1f in the main text). The second half-wave plate of $M_{\lambda/2}\left(\frac{\theta}{2}\right)$ is required to set the polarization close to the balance point, $\theta{\sim}45°$, for the detector. The Jones matrices for each optical element are expressed as,

$$M_{\lambda/2}(\theta) = e^{-i\frac{\pi}{2}} \begin{pmatrix} \cos^2\theta - \sin^2\theta & 2\cos\theta\,\sin\theta \\ 2\cos\theta\,\sin\theta & \sin^2\theta - \cos^2\theta \end{pmatrix} \qquad (S3)$$

$$M_{\lambda/4}(\theta) = e^{-i\frac{\pi}{4}} \begin{pmatrix} \cos^2\theta + i\sin^2\theta & (1-i)\cos\theta\sin\theta \\ (1-i)\cos\theta\sin\theta & \sin^2\theta + i\cos^2\theta \end{pmatrix} \tag{S4}$$

$$R = \begin{pmatrix} r_{xx} & r_{xy} \\ r_{yx} & r_{yy} \end{pmatrix}, \tag{S5}$$

where $r_{ii}$ is the reflection coefficient, which is related with the diagonal dielectric tensor component as $r_{ii} = \frac{1-\sqrt{\varepsilon_{ii}}}{1+\sqrt{\varepsilon_{ii}}}$, and $r_{ij}$ is related to the off-diagonal dielectric tensor components. A balanced detection with an application of electric current can be described as,

$$\Delta\theta(j_c) = \frac{1}{2}\frac{|E_{out}^x(j_c)|^2 - |E_{out}^y(j_c)|^2}{|E_{out}^x(j_c)|^2 + |E_{out}^y(j_c)|^2} = \frac{\Delta I(j_c)}{2I_0}. \tag{S6}$$

Finally, the lock-in detects the difference between $\Delta I(j)$ and $\Delta I(j = 0)$ with the current alternates between $j$ and $0$ with a modulation frequency of 3 kHz.

$$I_{LIA}(j) = \Delta I(j) - \Delta I(j = 0). \tag{S7}$$

In our experiment, the dielectric tensor can acquire contributions to its off-diagonal and diagonal elements. In a case, when the charge current affects the off-diagonal part of the dielectric tensor, the dielectric tensor modulation takes the following form:

$$\bar{\bar{\varepsilon}}(j_c = 0) = \begin{pmatrix} \varepsilon_{xx} & 0 \\ 0 & \varepsilon_{yy} \end{pmatrix} \rightarrow \bar{\bar{\varepsilon}}(j_c) = \begin{pmatrix} \varepsilon_{xx} & \delta_{xy} \\ -\delta_{xy} & \varepsilon_{yy} \end{pmatrix}, \tag{S8}$$

where the off-diagonal element $\delta_{xy}$ can be generated, for example, by the current-induced magnetization such as the valley Hall effect[9,10]. Then the complex angle $\Delta\theta$, plotted as a function of $\theta$, exhibits a $\cos(2\theta)$ behavior with a constant offset (Fig. S6a).

When the charge current modifies the diagonal part of the dielectric tensor, then the dielectric tensor modulation can be represented as follows:

$$\bar{\bar{\varepsilon}}(j_c = 0) = \begin{pmatrix} \varepsilon_{xx} & 0 \\ 0 & \varepsilon_{yy} \end{pmatrix} \rightarrow \bar{\bar{\varepsilon}}(j_c) = \begin{pmatrix} \varepsilon_{xx} + \delta_{xx} & 0 \\ 0 & \varepsilon_{yy} + \delta_{yy} \end{pmatrix}, \tag{S9}$$

where the corrections $\delta_{xx}$ and $\delta_{yy}$ could be driven by the charge carrier accumulation. In this case, the complex angle shift $\Delta\theta$ shows a distinctive $\sin(2\theta)$ behavior with the maximum signal at appearing, without any offset, at $\theta = 45°$ (Fig. S6b). The periodic $\sin(2\theta)$ behavior is produced due to the anisotropy between $\varepsilon_{xx}$ and $\varepsilon_{yy}$.

The dependence of $\Delta\theta$ on the anisotropy ratio $r = \frac{\varepsilon_{yy}}{\varepsilon_{xx}}$ is shown in Fig. S6c. There is no rotation of the polarization angle, $\Delta\theta = 0$, for an isotropic environment, $r = 1$. If the system is strongly anisotropic, then the signal follows a $\sin(2\theta)$ behavior, and the maxima are reached at two values of the angle, $\theta = 45°$ and $\theta = 135°$.

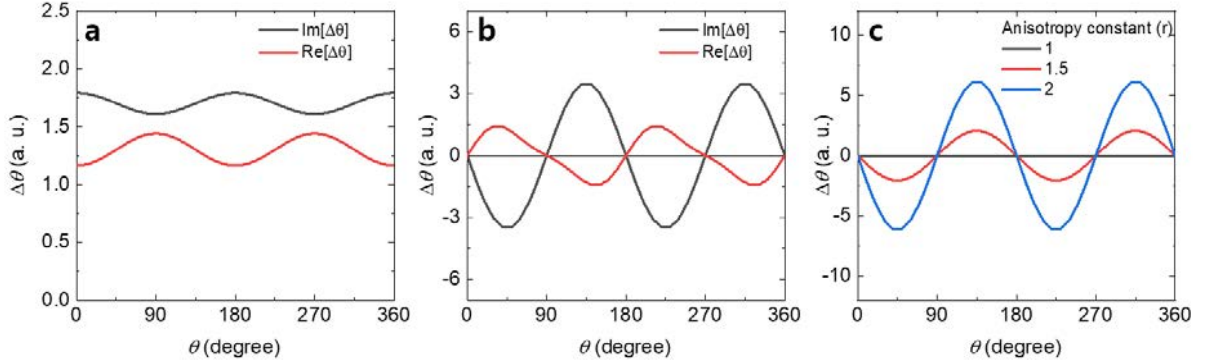

**Figure S6. a**, Rotation of polarization $\Delta\theta$ as the function of the azimuthal angle ($\theta$) in the case when the current induces the change in the off-diagonal permittivity, $\Delta\varepsilon_{ij} \neq 0$ ($i \neq j$). **b**, The same as in a, but for the diagonal permittivity, $\Delta\varepsilon_{ii} \neq 0$. **c**, Dependence of the polarization rotation $\Delta\theta$ on the anisotropic ratio $r$ in the case when the diagonal permittivity is modified, $\Delta\varepsilon_{ii} \neq 0$. The $\sin(2\theta)$ behavior emerges for a large anisotropy, $r > 1$, with the maximum absolute values of $\Delta\theta$ occurring at $\theta = 45°$, $135°$, $225°$, and $315°$.

## S6. Determination of the charge relaxation length

The relaxation length can be obtained by adopting the convolution procedure with the bi-exponential function:

$$\Delta\theta(x) \propto \int G(x-t)\Delta x(t)dt, \qquad (S10)$$

where $G(x)$ is the Gaussian function $\exp\left[-\left(\frac{x}{w_0}\right)^2\right]$ with the width $w_0$ of $\approx 1\ \mu m$ and $\Delta\theta(x)$ is the measured signal distribution along $x$-axis obtained with the Gaussian-shaped laser beam. We assume that the signal profile $\Delta\theta(t)$ has a $\sinh\left(\frac{x}{\lambda_a}\right)$ profile as predicted by the theory (SM S10). Figure S7 shows the convolution lines obtained for a number of values of the relaxation length $\lambda_a$ taken within the interval $0.2\ \mu m$ to $2\ \mu m$ with the step of $\Delta\lambda = 0.2\ \mu m$. A comparison of these convolution lines with the experimental data, shown in the same figure, makes it evident that the fit of the experimental data with the convolution function (S10) allows us to resolve a sub-micrometer decay length of the experimental data $\Delta\theta(x)$ with the use of the much wider, $\approx 1\ \mu m$, laser spot.

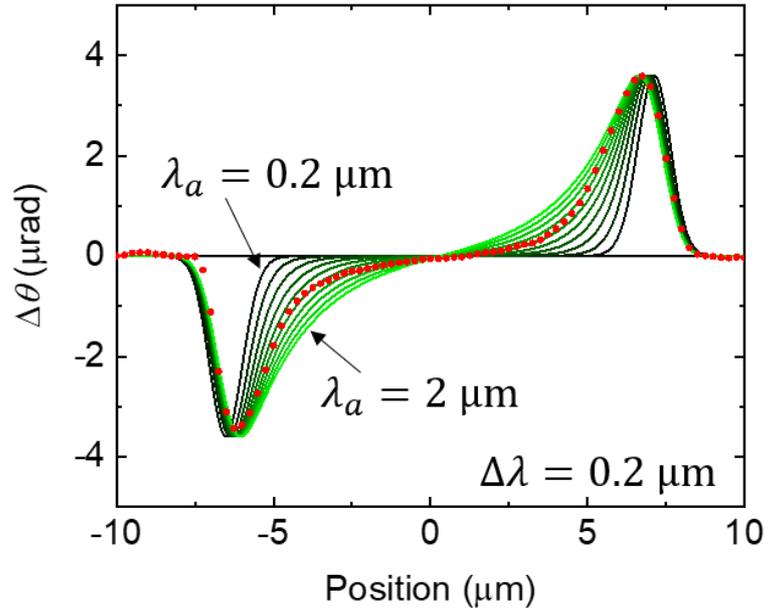

**Figure S7**. A relaxation profile data and the Gaussian-convoluted fitting lines. The red dots represent an example of experimental data along the *a*-axis of the device. The fitting lines are normalized to have the same maximum Kerr angle. The relaxation length $\lambda_a$ is varied in the interval from 0.2 $\mu$m to 2 $\mu$m with the step $\Delta\lambda = 0.2\,\mu$m. Using a fitting procedure, the relaxation length $\approx 1.4\,\mu$m can be resolved with an acceptable accuracy with the use of the beam of the width $\approx 1\,\mu$m.

## S7. Crystal axis dependence of the charge relaxation length

To study the effects of the anisotropy of Wte₂, we conduct 2D spatial mappings for the charge accumulation of the devices where the electrical current is injected separately in the *a*- and *b*-axes of the same flake as shown in Fig. S8. Importantly, the charge accumulation is observed at both axes, but with different screening lengths (Fig. S8a and Fig. S8d). Figures S8b and S8f show the line profiles of the longitudinal charge distributions along the *a*-axis (the red dashed line in Fig. S8a) and *b*-axis (the blue dashed line in Fig. S8d). The charge accumulation and its relaxation profile can be expressed as $\sinh\left(\frac{x}{\lambda}\right)$, where $x$ is a longitudinal position, and $\lambda$ is a relaxation length. We fit the data taking into account the effect of finite probe size. We calculate the convolution integral between the charge profile and a Gaussian function, $\exp\left[-\left(\frac{x}{w_0}\right)^2\right]$, where $w_0 \approx 1\,\mu$m is the $1/e^2$ beam waist of probe light (SM S6). Finally, we obtain the extremely long charge relaxation lengths, $\lambda_a = 1.4\,\mu$m and $\lambda_b = 0.7\,\mu$m for *a*-axis (Fig. S8b) and *b*-axis (Fig. S8f) directions, respectively. These lengths appear to be anomalously long in view of the fact that the WTe₂ semimetal possesses a high carrier concentration, $\sim 10^{20}$ cm$^{-3}$ (ref. 23 in the main text), which should correspond to the expected Debye screening length of less than 1 nm [Ref. 8].

We note that some devices exhibit similar long-ranged $\Delta\theta$ signals not only in the longitudinal direction along the current flow but also in the transverse edges with respect to the current direction, as shown in Fig. S8c (the scanning result along the blue dashed line of Fig. S8a) and Fig. S8e (the scanning result along the red dashed line of Fig. S8c). Regardless of the direction of the current, the relaxation lengths along each axis remain the same: $\lambda_a = 1.4\ \mu m$ (shown in Figs. S8b,e) and $\lambda_b = 0.7\ \mu m$ (shown in Fig. S8c,f). In the absence of the background magnetic field, we suspect that the transverse charge accumulation is due to a small mismatch between the current direction and the crystalline axis[11,12]. According to our theoretical arguments presented in Section S11, the mismatch in an electrically anisotropic crystal can induce a charge accumulation in the transverse directions with respect to the direction of the current.

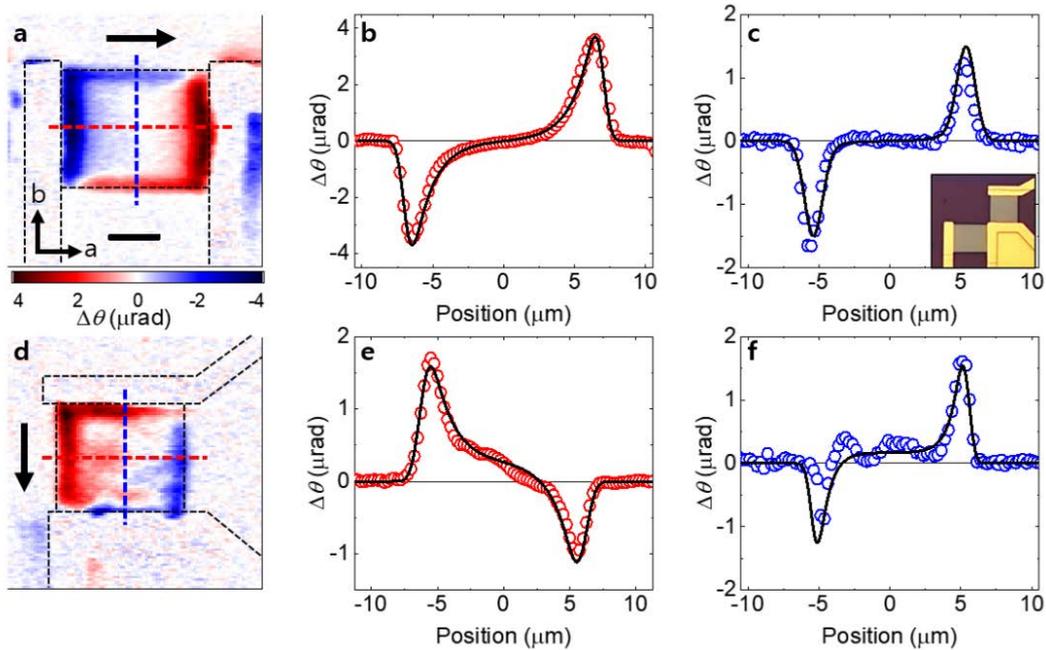

**Figure S8. Charge accumulation along different crystal directions of WTe₂. a**, **d**, The polarization rotation ($\Delta\theta$) mappings with the electric current of 2 mA along the *a*-axis and *b*-axis of WTe₂. The red and blue colors indicate the sign and the magnitude of $\Delta\theta$. The black scale bar in (a) corresponds to 5 $\mu$m. The black arrows in (a) and (d) indicate the direction of the current flow. **b**, **e**, The profile of $\Delta\theta$ along the *a*-axis corresponding to the red dashed lines in (a) and (d). **c**, **f**, The same the *b*-axis denoted by the blue dashed lines in (a) and (d). The open circles are the experimental data, while the black lines are the fits that use the convolution between the charge distribution of $\sinh(x/\tau)$ and probe-intensity profile of $\exp[-(x/w_0)^2]$. The fitted $\tau$ values are 1.4 and 0.7 $\mu$m for $a$- and $b$-axes, respectively. The inset in (c) shows an optical microscopy image of the investigated device.

# Supplementary Material (Theory)

In this theoretical section of the Supplementary Material, we first review known transport and charge-accumulation properties in two-component electron-hole systems. Then we present new calculations which support the pseudo-hydrodynamic interpretation of the charge-accumulation patterns observed experimentally.

## S8. TRANSPORT IN ELECTRON-HOLE SYSTEM

### A. Basic equations

In our considerations below, we do not advocate any particular model of the electronic band structure. Instead, we use a remarkably transparent and simple parabolic two-band model used in Refs. [13–15] which we follow closely in this introductory subsection. This model has a rather generic character.

We consider a generic two-component, electron-hole system which stays slightly away from charge neutrality point and possesses a weakly broken electron-hole symmetry. In a steady-state regime, the electronic ($\alpha = e$) and hole ($\alpha = h$) currents $\boldsymbol{j}_\alpha$, and the local fluctuations of the respective charge densities $\delta n_\alpha(\boldsymbol{r}) = n_\alpha(\boldsymbol{r}) - n_{0,\alpha}$ with respect to the constant background densities $n_{0,\alpha}$ are described, in the background of electric $\boldsymbol{E}$ and magnetic $\boldsymbol{B}$ fields, by the following macroscopic equations:

$$D_\alpha \boldsymbol{\nabla} \delta n_\alpha - \text{sign}(e_\alpha)\sigma_\alpha \boldsymbol{E} - \tau_\alpha \boldsymbol{j}_\alpha \times \boldsymbol{\omega}_\alpha + \boldsymbol{F}_\alpha = -\boldsymbol{j}_\alpha, \tag{S14}$$

$$\boldsymbol{\nabla} \cdot \boldsymbol{j}_\alpha = -\frac{1}{2}(\Gamma_e \delta n_e + \Gamma_h \delta n_h), \tag{S15}$$

where

$$\boldsymbol{F}_e = +\boldsymbol{F}_{eh}, \quad \boldsymbol{F}_h = -\boldsymbol{F}_{eh}, \quad \boldsymbol{F}_{eh} = \frac{\chi}{2}(\boldsymbol{j}_e - \boldsymbol{j}_h), \tag{S16}$$

is the friction force between the flows of electrons and holes controlled by the small parameter $\chi$. The electron and hole currents are proportional to the corresponding drift velocities $\boldsymbol{u}_\alpha$:

$$\boldsymbol{j}_\alpha = \frac{m\langle \boldsymbol{v}^2 \rangle \boldsymbol{u}_\alpha}{2}. \tag{S17}$$

The electric charges of holes and electrons are, respectively, $e_e = -e$ and $e_h = +e$, where $e = +|e|$ is the elementary electric charge.

The transport equations (S14) involve the diffusion coefficients

$$D_\alpha = \frac{\langle \boldsymbol{v}^2 \rangle_\alpha \tau_\alpha}{2\nu_{0,\alpha}}, \tag{S18}$$

the appropriate conductivities,

$$\sigma_\alpha = \frac{e n_{0,\alpha} \tau_\alpha}{m_\alpha} > 0, \tag{S19}$$

and the cyclotron frequencies

$$\boldsymbol{\omega}_\alpha = \frac{e_\alpha \boldsymbol{B}}{m_\alpha c}, \tag{S20}$$

which are expressed in terms of the masses $m_\alpha$, the momentum relaxation times $\tau_\alpha$, the background charge densities $n_{0,\alpha}$, and the electric charges of electrons ($e_e = -e$) and holes ($e_h = +e$) calculated in units of the fundamental electric charge $e = +|e|$.

The quantities $\nu_{0,\alpha}$ in Eq. (S18) are the proportionality coefficients (the susceptibilities) which relate the fluctuations of the local densities $\delta n_\alpha$ and the corresponding chemical potentials $\delta \mu_\alpha$:

$$\delta n_\alpha(\boldsymbol{r}) = \nu_{0,\alpha}\, \delta \mu_\alpha(\boldsymbol{r}). \tag{S21}$$

The recombination rates that enter the current conservation equations (S15),

$$\Gamma_e = 2\gamma n_{0,h}, \qquad \Gamma_h = 2\gamma n_{0,e}, \tag{S22}$$

are expressed via a model-dependent coefficient $\gamma$ which describes the electron-hole recombination process.

To avoid unnecessary complications that do not affect the qualitative features of the system, we set the equal masses for electrons and holes $m_h = m_e = m$, so that the cyclotron frequencies (S20) get simplified:

$$\boldsymbol{\omega}_e = -\boldsymbol{\omega}, \qquad \boldsymbol{\omega}_h = \boldsymbol{\omega}, \qquad \boldsymbol{\omega} = \frac{e \boldsymbol{B}}{mc}. \tag{S23}$$

It is also convenient to work at the exact charge-neutrality point in terms of electron and hole populations, $n_{0,e} = n_{0,h} = n_0$. This condition simplifies calculations without restricting the generality of our results. Furthermore, we take into account that the electron and hole recombination rates are the same (S22):

$$\Gamma_e = \Gamma_h = \frac{1}{\tau_R}, \tag{S24}$$

where we denoted by $\tau_R$ the common electron-hole recombination time.

It is convenient to introduce the charged current $\boldsymbol{j}$ and the neutral current of the quasi-particles $\boldsymbol{P}$:

$$\boldsymbol{j} = \boldsymbol{j}_e - \boldsymbol{j}_h, \qquad \boldsymbol{P} = \boldsymbol{j}_e + \boldsymbol{j}_h, \tag{S25}$$

and the charge and quasiparticle density fluctuations, respectively:

$$\delta n = \delta n_e - \delta n_h, \qquad \delta \rho = \delta n_e + \delta n_h. \tag{S26}$$

The electric current and the electric charge are defined, respectively, as follows:

$$\boldsymbol{J} = -e\boldsymbol{j}, \qquad \delta Q = -e\delta n. \tag{S27}$$

In order to discriminate between $\boldsymbol{J}$ and $\boldsymbol{j}$, we will call them the electric current and the charged current, respectively (and likewise for the densities $\delta Q$ and $\delta n$).

The linear combinations of Eqs. (S14) for electrons and holes give us the equations for charged and neutral currents and their densities in the disorder-dominated regime:

$$D_+ \boldsymbol{\nabla}\delta n + D_- \boldsymbol{\nabla}\delta \rho + \sigma_+ \boldsymbol{E} + (\tau_- \boldsymbol{j} + \tau_+ \boldsymbol{P}) \times \boldsymbol{\omega} = -(1+\chi)\boldsymbol{j}, \tag{S28a}$$

$$D_- \boldsymbol{\nabla}\delta n + D_+ \boldsymbol{\nabla}\delta \rho + \sigma_- \boldsymbol{E} + (\tau_+ \boldsymbol{j} + \tau_- \boldsymbol{P}) \times \boldsymbol{\omega} = -\boldsymbol{P}, \tag{S28b}$$

where we used the following combinations for the conductivities, momentum relaxation times and the diffusion coefficients, respectively [13, 14]:

$$\sigma_\pm = \sigma_e \pm \sigma_h \equiv \frac{2en_0}{m}\tau_\pm, \tag{S29}$$

$$\tau_\pm = \frac{1}{2}\left(\tau_e \pm \tau_h\right), \tag{S30}$$

$$D_\pm = \frac{1}{2}\left(D_e \pm D_h\right). \tag{S31}$$

We assume a (small) difference in the momentum relaxation times and the diffusion coefficients for electrons and holes, so that none of the quantities (S29), (S30) and (S31) is vanishing.

The transport relations (S28) should be supplemented by the continuity condition for the electric current,

$$\boldsymbol{\nabla} \cdot \boldsymbol{j} = 0, \tag{S32}$$

the relaxation for the neutral particle current,

$$\boldsymbol{\nabla} \cdot \boldsymbol{P} = -\frac{\delta \rho}{\tau_R}, \tag{S33}$$

and the Maxwell equation:

$$\boldsymbol{\nabla} \cdot \boldsymbol{E} = -4\pi e\delta n. \tag{S34}$$

The relaxation law (S33) follows from Eqs. (S15) and (S24).

The presence of (nearly) compensated charge carriers produces profound effects on transport properties in confined geometries [14]. For example, in the Hall experiment, in the regime of perfect electron-hole symmetry, the Hall currents of electrons and holes propagating in opposite directions cancel each other completely. As a result, the lateral voltage drop disappears. If the electron-hole symmetry is violated, then the classical Hall effects generated by electrons and holes do not cancel each other, and the Hall voltage is formed. Our observations show that for perfectly oriented crystals of WTe$_2$ at room temperature, the classical Hall current is relatively small, thus indicating an approximate validity of the electron-hole symmetry. A slight deviation from charge neutrality allows us to explain the observed charge accumulation pattern and nonlocal conductivity of the system.

In our paper, we are interested in the charge and quasiparticle accumulation effects in the absence of the background magnetic field. Since the Oersted magnetic field induced by the electric current is negligibly small, we set the cyclotron frequencies (S20) to zero for the rest of our considerations. In addition, we exclude the thermal transport from our considerations neglecting the local temperature fluctuations $\delta T(\boldsymbol{r}) = 0$, which should play a subleading role. In the disorder-dominated regime, the thermalization between the electronic subsystem (which includes both hole and electron flows) and the crystal lattice is much faster than the electron-hole recombination time, $\tau_{\mathrm{ph}} \ll \tau_R$ implying that the charge accumulation effects emerge in the already locally thermalized system. To simplify our calculations, we do not consider slight friction between electron and hole flows (S16) which does not bring any qualitatively new effects.

### B. Length scales in the absence of magnetic field

By setting a vanishing magnetic field $\boldsymbol{B} = 0$ and zero electron-hole friction $\chi = 0$, we bring the transport equations (S28) to the symmetric form:

$$D_+ \boldsymbol{\nabla} \delta n + D_- \boldsymbol{\nabla} \delta \rho + \sigma_+ \boldsymbol{E} = -\boldsymbol{j}, \tag{S35a}$$

$$D_- \boldsymbol{\nabla} \delta n + D_+ \boldsymbol{\nabla} \delta \rho + \sigma_- \boldsymbol{E} = -\boldsymbol{P}, \tag{S35b}$$

which can also be derived from Eq. (1) of the main text.

Applying the divergence operator to these equations and using Eqs. (S32), (S33), and (S34), we get two coupled second-order differential equations for charged and quasiparticle densities:

$$D_+ \left( \Delta - \lambda_+^{-2} \right) \delta n + D_- \Delta \delta \rho = 0, \tag{S36a}$$

$$D_- \left( \Delta - \lambda_-^{-2} \right) \delta n + D_+ \left( \Delta - \lambda_R^{-2} \right) \delta \rho = 0, \tag{S36b}$$

where

$$\lambda_\pm^2 = \frac{D_\pm}{4\pi e \sigma_\pm} \equiv \frac{1}{8\pi e} \frac{D_e \pm D_h}{\sigma_e \pm \sigma_h}, \tag{S37}$$

are two (squared) lengths expressed via the combinations of the electron and hole parameters,

and

$$\lambda_R = \sqrt{D_+ \tau_R} \equiv \sqrt{\frac{(D_e + D_h)\tau_R}{2}}, \tag{S38}$$

is an effective length associated with the electron-hole recombinations. Notice that the parameter $\lambda_-^2$ in Eq. (S37) can take negative values.

Excluding the quasiparticle density $\delta\rho$ from Eqs. (S36), we get the following equivalent equations for the charge density $\delta n$ and the quasiparticle density $\delta\rho$:

$$\left(\Delta^2 - \frac{1}{\xi_1^2}\Delta + \frac{1}{\xi_2^4}\right)\delta n = \left(\Delta^2 - \frac{1}{\xi_1^2}\Delta + \frac{1}{\xi_2^4}\right)\delta\rho = 0, \tag{S39}$$

where

$$\begin{aligned}
\xi_1 &= \left(\frac{D_+^2 - D_-^2}{D_+^2 \lambda_+^{-2} - D_-^2 \lambda_-^{-2} + D_+^2 \lambda_R^{-2}}\right)^{\frac{1}{2}} = \left[4\pi e\left(\frac{\sigma_e}{D_h} + \frac{\sigma_h}{D_e}\right) + \frac{D_e + D_h}{2\tau_R D_e D_h}\right]^{-1/2} \\
&\equiv \left[\lambda_{\mathrm{TF}}^{-2} + \frac{1}{4}\left(\sqrt{\frac{D_e}{D_h}} + \sqrt{\frac{D_h}{D_e}}\right)^2 \lambda_R^{-2}\right]^{-1/2}, \tag{S40a}
\end{aligned}$$

$$\begin{aligned}
\xi_2 &= \left(\frac{D_+^2 - D_-^2}{D_+^2}\right)^{\frac{1}{4}} \sqrt{\lambda_+ \lambda_R} = \left(\frac{D_e D_h \tau_R}{4\pi e(\sigma_e + \sigma_h)}\right)^{\frac{1}{4}} \\
&\equiv \sqrt{2}\left(\sqrt{\frac{D_e}{D_h}} + \sqrt{\frac{D_h}{D_e}}\right)^{-1/2} \sqrt{\lambda_+ \lambda_R}, \tag{S40b}
\end{aligned}$$

are the two new length scales which control the spatial variation of the charge density $\delta n$ in the sample. Notice that both lengths (S40) are nonvanishing, real-valued parameters.

The first screening length (S40a) is expressed via the recombination length (S38) as well as the Thomas-Fermi length:

$$\lambda_{\mathrm{TF}} = \sqrt{\frac{D_e D_h}{4\pi e(\sigma_e D_h + \sigma_h D_e)}}. \tag{S41}$$

In the illustrative example when the electron/hole average velocities and the corresponding susceptibilities satisfy the relation $\langle \boldsymbol{v}^2 \rangle_e / \nu_{0,e} = \langle \boldsymbol{v}^2 \rangle_h / \nu_{0,h}$, the electron and hole diffusivities (S18) are proportional to the corresponding relaxation times, $D_e/\tau_e = D_h/\tau_h$, implying, together with Eq. (S29), that $\lambda_+ = \lambda_- = \lambda_{\mathrm{TF}}$, were the Thomas-Fermi length takes the familiar form:

$$\lambda_{\mathrm{TF}} = \left(\frac{1}{4\pi e^2 n_0 \nu_0} \frac{m\langle \boldsymbol{v}^2 \rangle}{2}\right)^{\frac{1}{2}}. \tag{S42}$$

This particular situation can be realized when the electron-hole symmetry is almost exact, with the only difference in the relaxation times, $\tau_e \neq \tau_h$.

What is the hierarchy of the length scales (S40) close to the neutrality point? The Thomas-Fermi length, given by Eq. (S41) or Eq. (S42), determines the distance scale of the electrostatic screening. In typical metals, the charge screening length is very short, of the order of a few nanometers while the e-h recombination length varies from hundreds of nanometers to centimeters depending on material and temperature [13]. Given the large difference in these lengths,

$$\lambda_R \gg \lambda_{\mathrm{TF}} \,. \tag{S43}$$

we can neglect the relaxation length in Eq. (S40a) and assume that the electron and hole relaxation times are close to each other, $\tau_e \simeq \tau_h$ with $|\tau_e - \tau_h| \ll \tau_e + \tau_h$ which is a reasonable assumption close to the electron-hole symmetry point. One then gets:

$$\xi_1 \simeq \lambda_{\mathrm{TF}}, \qquad \xi_2 \simeq \sqrt{\lambda_{\mathrm{TF}} \lambda_R} \qquad (\tau_h \simeq \tau_e), \tag{S44}$$

implying that $\xi_2 \gg \xi_1$ due to the natural scale hierarchy of the e-h relaxation and charge screening lengths (S43).

It is convenient to rewrite Eq. (S39) in the form of Eq. (2) of the main text:

$$\left( \Delta - \varkappa_1^2 \right) \left( \Delta - \varkappa_2^2 \right) \delta n(\boldsymbol{r}) = 0, \tag{S45}$$

where the quantities

$$\varkappa_{1,2} \equiv \frac{1}{\lambda_{1,2}} = \frac{1}{\sqrt{2}} \left[ \frac{1}{\xi_1^2} \pm \left( \frac{1}{\xi_1^4} - \frac{4}{\xi_2^4} \right)^{\frac{1}{2}} \right]^{\frac{1}{2}} \tag{S46}$$

characterize two new inverse length scales $\lambda_1$ and $\lambda_2$ via the length parameters $\xi_1$ and $\xi_2$ given explicitly in Eq. (S40). In Eq. (S46), the signs "+" and "−" correspond, respectively, to the quantities $\varkappa_1$ and $\varkappa_2$ arranged according to the hierarchy:

$$\varkappa_1 > \varkappa_2 > 0 \,. \tag{S47}$$

In the near-neutrality regime with the natural difference between the length scales (S43), one gets:

$$\varkappa_1 = \frac{1}{\xi_1} \left[ 1 - \frac{1}{2} \left( \frac{\xi_1}{\xi_2} \right)^4 + O\left( \left( \frac{\xi_1}{\xi_2} \right)^8 \right) \right] \,, \tag{S48}$$

$$\varkappa_2 = \frac{\xi_1}{\xi_2^2} \left[ 1 + \frac{1}{2} \left( \frac{\xi_1}{\xi_2} \right)^4 + O\left( \left( \frac{\xi_1}{\xi_2} \right)^8 \right) \right] \,. \tag{S49}$$

Using Eq. (S44), we conclude that the quantities $\varkappa_1$ and $\varkappa_2$ are determined by the Thomas-Fermi length and the electron-hole recombination length, respectively:

$$\varkappa_1 \simeq \frac{1}{\lambda_{\mathrm{TF}}}, \qquad \varkappa_2 \simeq \frac{1}{\lambda_R}, \qquad \varkappa_1 \gg \varkappa_2 \,. \tag{S50}$$

## S9.  A LINK TO TWO-DIMENSIONAL (PSEUDO)HYDRODYNAMICS

### A.  Charge density in two-component systems and stream function

In electronic fluids, the hydrodynamic behavior of electrons naturally leads to negative values of the local (vicinity) resistance $R_v$. The effect appears due to the emergence of the electronic backflow caused by the whirlpools in a viscous electronic fluid that are created, for instance, near current-injecting contacts. The striking example of the nonlocal electrostatic response emerges in graphene [16–21]. The effect can be described by hydrodynamics of one-component electronic fluid with the local velocity $\boldsymbol{u} = \boldsymbol{u}(\boldsymbol{r})$ which satisfies the Navier-Stokes equation for a steady flow:

$$-\eta \Delta \boldsymbol{u} - \left( \zeta + \frac{\eta}{3} \right) \boldsymbol{\nabla} \left( \boldsymbol{\nabla} \cdot \boldsymbol{u} \right) = \boldsymbol{f} - \nu \boldsymbol{u}, \tag{S51}$$

where $\eta$ and $\zeta$ are the shear and bulk viscosities of the fluid, $P$ is the pressure and $n$ is the local electronic density. We do not consider the magnetic-field background and omit the terms nonlinear in the fluid velocity $\boldsymbol{u}$ since our experimental data indicates that the Oersted magnetic fields are negligible while the optical experiment implies the absence of nonlinearities because we do not observe higher-harmonic responses in our experiments.

The first term in the right-hand side of Eq. (S51) is the force exerted on the unit-volume fluid element:

$$\boldsymbol{f} = -\boldsymbol{\nabla} P - ne\boldsymbol{E}. \tag{S52}$$

It includes the gradient of pressure $P$ which can be considered, in general, as a consequence of an external force $\boldsymbol{f}_{\text{ext}} = -\boldsymbol{\nabla} P$ applied to the sample [16]. Following closely Ref. [16] in this subsection, we neglect the effects related to the pressure inhomogeneities by setting $\boldsymbol{\nabla} P = 0$.

The second term in Eq. (S52) is the electrostatic force exerted on electrons by the local electric field

$$\boldsymbol{E} = -\boldsymbol{\nabla} \phi, \tag{S53}$$

where $\phi$ is the electrostatic potential.

The last term in the right-hand-side of Eq. (S51) involves, via the parameter $\nu = mn/\tau_e$, the momentum-relaxation scattering time $\tau_e$ which characterizes the diffusive relaxation of the electronic momentum density $\boldsymbol{p} = mn\boldsymbol{u}$ due to the lattice disorder and interactions with the phonons [16]. Here the quantity $m$ plays a role of the effective mass of the particle. The validity of hydrodynamics requires that the momentum-conserving electron-electron scattering proceeds with shorter relaxation times $\tau_{\text{mc}}$ as compared to the momentum-relaxing disorder scattering, $\tau_{\text{mc}} \ll \tau_{\text{mr}}$.

The flow generates the local electric current density

$$\boldsymbol{J} = en\boldsymbol{u},\tag{S54}$$

subjected to the conservation equation:

$$\boldsymbol{\nabla} \cdot \boldsymbol{J}(\boldsymbol{r}) = 0\,.\tag{S55}$$

For the charged fluid velocities smaller than the plasmonic velocities, the electronic density can be safely considered as a spatially homogeneous quantity, $n(\boldsymbol{r}) = n_0$. The conservation of electric charge (S55) then implies that the electronic fluid is described by a divergenceless velocity field, $\boldsymbol{\nabla} \cdot \boldsymbol{u} = 0$ and corresponds to an incompressible flow. Consequently, the second term in the left-hand side of the Navier-Stokes equation (S51) vanishes, and the bulk viscosity does not play any role.

With all these simplifications, the Navier-Stokes equation (S51) reduces to the following differential equation [16, 17]:

$$\sigma_0 \boldsymbol{\nabla}\phi(\boldsymbol{r}) + D_v^2 \Delta \boldsymbol{J}(\boldsymbol{r}) - \boldsymbol{J}(\boldsymbol{r}) = 0,\tag{S56}$$

where

$$\sigma_0 = \frac{n_0 e^2}{m}\tau_e\,,\tag{S57}$$

is the Drude-like conductivity for the particles with electric charge $e$, density $n_0$, and mass $m$ which scatter over each other with the momentum-relaxing time $\tau_e$. Equation (S56) corresponds to Eq. (4) of the main text.

The quantity $D_v$ has the dimension of length and plays the role of a diffusion constant. In the absence of viscous behavior, $\eta = 0$, the diffusion constant vanishes, $D_v = 0$, and the Navier-Stokes equation (S56) reduces to the Ohmic flow $\boldsymbol{J} = \sigma_0 \boldsymbol{E}$ with the conductivity (S57).

It is convenient to describe a two-dimensional incompressible flow via the stream function $\psi = \psi(\boldsymbol{r})$,

$$\boldsymbol{u} = \hat{\boldsymbol{z}} \times \boldsymbol{\nabla}\psi,\tag{S58}$$

where $\hat{\boldsymbol{z}}$ is a unit out-of-plane vector. In the components, $u_x = -\partial_y\psi$ and $u_y = \partial_x\psi$, implying that the vorticity of the incompressible flow (S58) takes the simple form: $\boldsymbol{\nabla} \times \boldsymbol{u} = \hat{\boldsymbol{z}}\,\Delta\psi$. Applying the curl operator to the Navier-Stokes equation (S56) and using Eq. (S54), we then get the following equation for the stream function [17]:

$$\Delta\left(\Delta - \frac{1}{D_v^2}\right)\psi = 0,\tag{S59}$$

which is also given in Eq. (3) of the main text.

The Navier-Stokes equation (S59) for the stream function $\psi$ in the single-component electronic system is strikingly similar to Eq. (S39) which describes the behavior of the charge density $\delta n$ in the two-component, electron-hole system close to the neutrality point. The role of the stream function $\psi$ is played by the charge density $\delta n$, while the diffusion length $D_v$ is associated with the first length scale $\xi_1$ given in Eq. (S40a). Using the association of the length scales (S50), we notice that at for the lengths of the order of the shortest length in the system, $|\boldsymbol{x}| \sim \lambda_{\mathrm{TF}} \ll \lambda_R$, the equation for the particle density (S45) resembles the hydrodynamic equation (S59) for the stream function with

$$\Delta \left( \Delta - \lambda_{\mathrm{TF}}^{-2} \right) \delta n(\boldsymbol{r}) \simeq 0. \tag{S57}$$

The particle-hole recombination is not efficient at short length scales, and the number of quasiparticles is conserved. In this limit, Eq. (S60) acquires a formal mathematical resemblance with the hydrodynamic relation (S59).

While this curious association may seem to have a very formal character, we show below that the neutral quasiparticle current $\boldsymbol{P}$ develops large whirlpool-like structures in the bulk of the sample and the backflow in close vicinity of the current-injecting contacts. These structures are responsible for the experimentally observed sign-changing charge accumulation pattern discussed in our paper. The reason is that the high-derivative nature of Eq. (S60) allows for the appearance of the non-harmonic solutions, which lead to the emergence of a non-Ohmic component of the *neutral* quasiparticle current $\boldsymbol{P}$. As we show below, the latter, non-harmonic component of the *neutral* quasiparticle flow $\boldsymbol{P}$ is responsible for the sign-alternating charge accumulation pattern of *electric charge* density $\delta n$ observed near the contacts. Moreover, the very same, non-harmonic property of the system leads to the sign-alternating pattern of electrostatic potential observed in graphene [16]. Thus, the analogy of Eq. (S60) with the hydrodynamic relation (S59) is closer than one might seem from first sight.

Finally, we notice that similarly to the electric current $\boldsymbol{j}$, the quasiparticle current $\boldsymbol{P}$ possesses zero curl, $\boldsymbol{\nabla} \times \boldsymbol{P} \equiv 0$, because $\boldsymbol{P}$ is a pure gradient flow according to Eqs. (S35b) and (S53). The quasiparticle flow velocity $\boldsymbol{v} = \boldsymbol{P}/\delta\rho$ can, however, can exhibit signatures of nonzero vorticity, $\boldsymbol{\nabla} \times \boldsymbol{v} = (\boldsymbol{P} \times \boldsymbol{\nabla}\delta\rho)/\delta\rho \neq 0$, because the quasiparticle current $\boldsymbol{P}$ represents a compressible flow and the local quasiparticle density $\delta\rho = -\tau_R \boldsymbol{\nabla} \cdot \boldsymbol{P}$ is not a globally uniform quantity.

## B. Navier-Stockes equation with higher-order Burnett terms

The two-component system of electrons and holes (S35) possesses two types of currents: the conserved (S32) electric current $\boldsymbol{J}$ and the non-conserved (S33) neutral particle current $\boldsymbol{P}$. The electric current $\boldsymbol{J}$ represents a non-hydrodynamic flow which is described, as we will see below, by a coordinate-independent quantity in one spatial dimension and by a harmonic function in two dimensions. In the regimes considered in this article, the charged current does not exhibit a backflow and cannot lead, for example, to a negative resistance, similar to the effect found in graphene.

Curiously enough, the neutral current $\boldsymbol{P}$, given by a chargeless sum of the electrons and holes, can be interpreted as a hydrodynamic flow of compressible fluid described by a Navier-Stokes equation equipped with higher-order derivatives and generalized forces. In order to demonstrate this property, we apply the operator $D_+ \left( \Delta - \lambda_+^{-2} \right)$ to Eq. (S35b) and use Eq. (S36a) to express the charged particle density $\delta n$ via the neutral density $\delta \rho$. Next, we use the relaxation of the neutral charge (S33) to represent the neutral particle density $\delta n$ via the neutral current $\boldsymbol{P}$. We get the following equation for the neutral current:

$$-\frac{\lambda_+^2}{\tau_R} \Delta \boldsymbol{P} - D_+ \boldsymbol{\nabla} (\boldsymbol{\nabla} \cdot \boldsymbol{P}) + \frac{D_e D_h}{4\pi e \sigma_+} \Delta \boldsymbol{\nabla} (\boldsymbol{\nabla} \cdot \boldsymbol{P}) = -\frac{1}{\tau_R} \boldsymbol{P} - \frac{\sigma_-}{\tau_R} \left( \lambda_+^2 \Delta - 1 \right) \boldsymbol{E} . \quad \text{(S61)}$$

Despite its cumbersome form, this equation resembles quite closely the linearized Navier-Stokes equation (S51):

- The first two terms at the left-hand side of these two equations have the same differential form if one associates the current $\boldsymbol{P}$ with the velocity $\boldsymbol{u}$ of a fluid [we can safely ignore the difference in dimensions of these two quantities given the linearity of Eq. (S61)].

- The first term in the right-hand side of Eq. (S61) perfectly matches the exact Navier-Stokes counterpart in Eq. (S51): it describes the momentum relaxation of a fluid volume element. The recombination time $\tau_R$ plays the role of the momentum-relaxation time $\tau_{\mathrm{mr}}$.

- The second term in the right-hand side depends only on the background electric field and thus can be associated with a generalized force $\boldsymbol{F}$ which also appears in the linearized Navier-Stokes equation (S51).

- The only principal difference between the differential equation (S61) that describes the neutral current $\boldsymbol{P}$ and the usual Navier-Stokes equation (S51) emerges in the third term of the left-hand side of Eq. (S61). This four-derivative term is absent in the standard Navier-Stokes equation, but it can appear in the higher-order generalizations of the hydrodynamic equations known as Burnett and super-Burnett equations [22]. However, due to the high order of derivatives, this term is irrelevant for long wavelengths and can be neglected in the infrared limit.

The lowest-order terms of the Navier-Stokes equation (S61) are represented in Eq. (5) of the main text.

Summarising, the neutral current $\boldsymbol{P}$ can now be interpreted as a new velocity variable, $\boldsymbol{u} \leftrightarrow \boldsymbol{P}$, of a compressible and non-conserved hydrodynamic flow. The equation for the neutral quasiparticle current (S61) can be rewritten in the suggestive hydrodynamical form:

$$-\bar{\eta} \Delta \boldsymbol{u} - \left( \bar{\zeta} + \frac{\bar{\eta}}{3} \right) \boldsymbol{\nabla} (\boldsymbol{\nabla} \cdot \boldsymbol{u}) - \Lambda \Delta \boldsymbol{\nabla} (\boldsymbol{\nabla} \cdot \boldsymbol{u}) = \boldsymbol{f}_{\mathrm{eff}} - \frac{1}{\tau_{\mathrm{mr}}} \boldsymbol{u}. \quad \text{(S62)}$$

The association of the hydrodynamical variables (the kinematic share viscosity $\bar{\eta}$, the kinematic bulk viscosity $\bar{\zeta}$, the Burnett coupling $\Lambda$, and the effective force acting on the neutral quasiparticle flow $\boldsymbol{f}_{\text{eff}}$) with the parameters of the electron-hole system (conductivities, diffusion constants, recombination time) is shown in Table I.

| shear viscosity | $\bar{\eta}$ | $\frac{\lambda_+^2}{\tau_R}$ |
|---|---|---|
| bulk viscosity | $\bar{\zeta}$ | $D_+ - \frac{\lambda_+^2}{3\tau_R}$ |
| (super)Burnett coupling | $\Lambda$ | $\frac{2 D_e D_h}{D_e + D_h} \lambda_+^2$ |
| generalized force | $\boldsymbol{f}_{\text{eff}}$ | $\frac{\sigma_-}{\tau_R} \left( 1 - \lambda_+^2 \Delta \right) \boldsymbol{E}$ |
| momentum-relaxation time | $\tau_{\text{mr}}$ | $\tau_R$ |
| fluid velocity | $\boldsymbol{u}$ | $\boldsymbol{P}$ |

TABLE I. The association of the quantities in the transport equation (S62) for the neutral quasiparticle current $\boldsymbol{P}$ with the hydrodynamical variables of the linearized Navier-Stokes equation (S61).

Thus, we demonstrated analytically that the neutral quasiparticle current $\boldsymbol{P}$ could be associated with a compressible fluid that obeys a linearized Navier-Stokes equation equipped with a higher-derivative velocity term and a generalized force acting on a fluid element. Moreover, in the long-wavelength limit, at the distances more extended than the Thomas-Fermi length ($\sim \lambda_+$), the higher-derivative (super)Burnett term and a higher-derivative correction to the generalized force drop out from the Navier-Stokes equation (S62) which then reduces to the standard linearized hydrodynamic equation for the neutral flow shown in Eq. (5) of the main text of our work.

Finalizing this subsection, we stress that the formal resemblance of the transport equation (S62) with the higher-order Navier-Stokes equation does not mean in any way that we have real hydrodynamics in our system for which the momentum-conserved length scale – that governs the collisions between electrons and holes – is shorter than the momentum-relaxing length (the latter quantity is given by mean scattering length of electrons or holes with impurities or phonons). While the hydrodynamic regime in WTe$_2$ is known to be realized around $20\,\text{K}$, it is excluded at room temperature [23] which we consider in our article.

On the other hand, we show theoretically that the Navier-Stokes-like form of the transport equation for the neutral current in the two-carrier systems (S62) leads to the hydrodynamic-like features marked by the appearance of the backflow and whirlpools. Moreover, while the neutral current itself cannot be observed in our experiments, we show that the neutral particle backflow produces the sing-alternating patterns of electric charge accumulation that we were able to detect experimentally in room-temperature WTe$_2$. To stress the hydrodynamic features in the non-hydrodynamic regime, we use the term "*pseudo-hydrodynamics*" concerning the two-carrier transport in WTe$_2$. In the passing, we notice that the pseudo-hydrodynamic regime appears because of the deviation from the charge neutrality regime and, simultaneously, due to the existence of the long recombination time. The latter property implies that the inelastic, momentum-relaxing recombination of electrons and holes proceeds slower than any other processes in the system, including the momentum-preserving kinetic equilibration within either electron and hole subsystems and the momentum-relaxing scattering over phonons and impurities.

## S10. CHARGE ACCUMULATION IN ONE-DIMENSIONAL CONDUCTOR

The accumulation of electric charge density and quasiparticle density naturally occurs in the background of the magnetic field in the Hall geometry [13, 14]. The effect appears due to the transverse voltage gradient caused by the Lorentz force acting on the drifting charge carriers. In this section of Supplementary Material, we discuss the features of the charge accumulation in the absence of a background magnetic field. The effect occurs in the longitudinal direction parallel to the Ohmic drift of the charge carriers.

### A. Generic one-dimensional solution

Before addressing the two-dimensional problems, it is instructive to consider the simplest one-dimensional example, which illustrates the basic properties of the two-carrier system close to the neutrality point. Despite its simplicity, even this one-dimensional example has some nontrivial properties, which are the precursors of a more complicated behavior that emerges in the two-dimensional case, which we will consider in detail later. A similar nontrivial one-dimensional behavior is also observed for the lateral charge accumulation in the Hall effect in two-component nearly-compensated conductors [13, 14]. As there are geometrical differences in the charge-accumulation mechanisms in our one-dimensional setup and the nearly-compensated Hall effect, we discuss the one-dimensional case in very detail.

Consider a wide uniform conductor with the length $L$ and width $W$ much larger than the recombination length, $L \gg \lambda_R$ and $W \gg \lambda_R$.

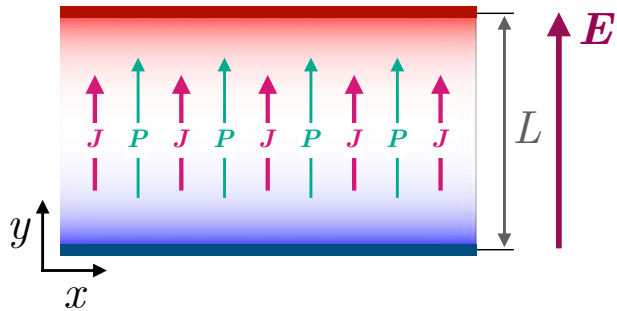

FIG. S8. One-dimensional character of the electric charge accumulation in a two-component conductor with wide electrodes attached to its upper and lower sides.

Given the linearity of the equations of motion, the transport in the middle of the sample features a one-dimensional current flow shown in Fig. S8. The charged current $\boldsymbol{j} = j\mathbf{e}_y$ is injected (absorbed) at the upper, $y = L/2$ (lower, $y = -L/2$) side of the slab. Both these currents,

$$\boldsymbol{j}(y)\bigg|_{y=\pm L/2} = (0, j), \tag{S63}$$

have no dependence on the coordinate $x$ along the boundary, so that the solution inside the sample has a one-dimensional character: $\delta\rho = \delta\rho(y)$, $\delta n = \delta n(y)$, etc. Notice that according to Eq. (S27), the electric current $J$ is related to the charged current $j$ via the relation $J = -ej$.

The continuity (S32) of the electric current $\boldsymbol{j}(x, y) = (0, j_y(y))$ implies $\partial_y j_y(y) = 0$. Therefore, the charged current is a constant quantity inside the sample $j_y(y) = j$, which coincides with the value that the current (S63) takes at each sides of the slab.

In one spatial dimension, the equation for the charge density (S45) reduces to the simple differential equation:

$$\left(\partial_y^2 - \varkappa_1^2\right)\left(\partial_y^2 - \varkappa_2^2\right)\delta n(y) = 0, \tag{S64}$$

with the general solution

$$\delta n(y) = \sum_{s=\pm}\sum_{a=1}^{2} n_a^s e^{s\varkappa_a y}, \tag{S65}$$

The solution is entirely characterized by four arbitrary parameters $n_{1,2}^{\pm}$ which will be fixed later via appropriate boundary conditions.

The electric field $\boldsymbol{E}(x, y) = (0, E_y(y))$ is related to the electric charge density (S65) with the help of the one-dimensional Maxwell equation (S34):

$$\partial_y E_y(y) = -4\pi e\,\delta n(y). \tag{S66}$$

Its solution,

$$E_y(y) = E_0 + \sum_{s=\pm}\sum_{a=1}^{2} e_a^s e^{s\varkappa_a y}, \tag{S67}$$

contains a homogeneous, coordinate-independent contribution $E_0$ supplemented by the inhomogeneous, space-dependent part. The latter is rigidly tightened to the charge density (S65), and its parameters $n_{1,2}^{\pm}$, via the four amplitudes:

$$e_a^{\pm} = \mp C_a^{(E)} n_a^{\pm}, \qquad a = 1, 2. \tag{S68}$$

where the proportionality coefficients are as follows:

$$C_a^{(E)} = \frac{4\pi e}{\varkappa_a} \,.$$ (S69)

In the near-neutrality regime,

$$C_1^{(E)} \ll C_2^{(E)} \,,$$ (S70)

because $\varkappa_1 \simeq \lambda_{\text{TF}}^{-1} \gg \lambda_R^{-1} \simeq \varkappa_2$. Notice that in the chosen near-neutrality regime,

$$C_1^{(\rho)} \simeq \frac{D_+}{D_-} \left( \frac{\lambda_{\text{TF}}}{\lambda_+^2} - 1 \right) \equiv \frac{\sigma_e D_h - \sigma_h D_e}{\sigma_e D_h + \sigma_h D_e} \,,$$ (S71)

$$C_2^{(\rho)} \simeq \frac{D_e + D_h}{D_e - D_h} \left( \frac{\lambda_R}{\lambda_+} \right)^2 \,.$$ (S72)

Since $\lambda_+ \sim \lambda_{\text{TF}} \gg \lambda_R$, then $\varkappa_1 \gg \varkappa_2$ (because $\varkappa_1 \simeq \lambda_{\text{TF}}^{-1}$ and $\varkappa_2 \simeq \lambda_R^{-1}$) which implies the following hierarchy:

$$C_1^{(\rho)} \ll C_2^{(\rho)} \,.$$ (S73)

In our regime,

$$C_1^{(P)} \simeq \frac{D_e D_h}{D_e - D_h} \left( \frac{\lambda_{\text{TF}}}{\lambda_R} \right)^4 \,,$$ (S74)

$$C_2^{(P)} \simeq \frac{2 D_e D_h}{D_e - D_h} \left( \frac{\lambda_R}{\lambda_{\text{TF}}} \right)^2 \,,$$ (S75)

and

$$C_1^{(P)} \ll C_2^{(P)} \,.$$ (S76)

The neutral and charged particle currents take the following form (S35):

$$D_+ \partial_y \delta n + D_- \partial_y \delta \rho + \sigma_+ E_y = -j \,,$$ (S77a)
$$D_- \partial_y \delta n + D_+ \partial_y \delta \rho + \sigma_- E_y = -P_y \,.$$ (S77b)

According to Eq. (S77a), the space-independent component $E_0$ of the electric field (S67) gives us the dissipative Ohm law for the charged current:

$$j = -\sigma_+ E_0 \,,$$ (S78)

In the conventional notations, this relation corresponds to the Ohm law for the electric

current:

$$J = -ej = \sigma E_0 \,, \tag{S79}$$

with the electric conductivity $\sigma \equiv e\sigma_+$.

The consistency of the solution requires that the homogeneous part of the electric field $E_0$ should be related to the charged current $j$ – which consists only of a homogeneous part – via the Ohm law (S78). This requirement implies the following relation for the electric field (S67) at the boundaries of the sample:

$$E_y(y)\bigg|_{y=\pm L/2} = E_0 \,. \tag{S80}$$

The very same equation (S77a), applied now to the inhomogeneous components of the densities $\delta n$ and $\delta\rho$, as to the electric field (S67), gives us the particle density:

$$\delta\rho(y) = \sum_{s=\pm} \sum_{a=1}^{2} \rho_a^s e^{s\varkappa_a y} \,, \tag{S81}$$

with the amplitudes:

$$\rho_a^\pm = C_a^{(\rho)} n_a^\pm \,, \tag{S82}$$

where the proportionality coefficients are as follows:

$$C_a^{(\rho)} = \frac{D_+}{D_-} \left( \frac{1}{\lambda_+^2 \varkappa_a^2} - 1 \right) \,. \tag{S83}$$

Finally, the last transport equation (S77b) determines the quasiparticle flow $\boldsymbol{P} = (0, P_y(y))$ through the sample:

$$P_y(y) = P_0 + \sum_{s=\pm} \sum_{a=1}^{2} s\varkappa_a p_a^\pm e^{\pm\varkappa_a y} \,. \tag{S84}$$

This quantity is expressed via the following amplitudes:

$$p_a^\pm = C_a^{(P)} n_a^\pm \,, \tag{S85}$$

where

$$C_a^{(P)} = \frac{1}{D_-} \left( D_+^2 - D_-^2 - \frac{D_+^2 \lambda_+^{-2} - D_-^2 \lambda_-^{-2}}{\varkappa_a^2} \right) \equiv \frac{2D_e D_h}{D_e - D_h} \left( \frac{1}{\lambda_{\text{TF}}^2 \varkappa_a^2} - 1 \right) \,. \tag{S86}$$

An analog of the Ohm law determines the homogeneous contribution to the neutral quasiparticle current (S84) for the neutral current (S78):

$$P_0 = -\sigma_- E_0 \,. \tag{S87}$$

Contrary to the electric current (S78), the uniform contribution to the neutral quasiparticle current (S87) naturally vanishes at the perfect neutrality point ($\sigma_- = \sigma_e - \sigma_h = 0$) due to the exact compensation of the electron and hole currents.

Finally, the relaxation law for the quasiparticle current (S33), expressed in one spatial dimension as $\partial_y P_y(y) = -\delta\rho(y)/\tau_R$, is automatically satisfied by virtue of Eqs. (S40), (S46) and (S77b).

The boundary condition determines the precise form of the solution on the electric field at the boundaries of the slab (S80) accompanied by the hard-wall condition for the neutral quasiparticle current. As a result, the quasiparticles cannot exit the sample, and therefore the normal component of the quasiparticle current vanishes at upper and lower electrodes of the sample, Fig. S8:

$$P_y(y)\bigg|_{y=\pm L/2} = 0\,. \tag{S88}$$

We remind that the only free (external) parameter of the system is the electric current $J \equiv -ej$. This parameter determines, via the Ohm law (S78), the homogeneous part of the electric field, which, in turn, fixes the homogeneous part of the neutral current (S88). Next, the condition for the neutral current (S87) relates the homogeneous part of the neutral current to the inhomogeneous component of the same neutral current; both are seen in Eq. (S84). This chain of relations determines the whole solution of the system since the amplitudes $n_{1,2}^\pm$, $e_{1,2}^\pm$, $\rho_{1,2}^\pm$, and $p_{1,2}^\pm$, that characterize the inhomogeneous parts of the appropriate quantities, are already linked to each other in Eqs. (S68), (S82), (S85).

We get the following result for the coefficients:

$$n_a^\pm = \pm(-1)^a \frac{\varkappa_a}{\beta_1\varkappa_1 - \beta_2\varkappa_2}\frac{\sigma_- J}{2e\sigma_+}\frac{1}{\cosh\frac{\varkappa_a L}{2}}\,, \tag{S89}$$

where we used the relation $j = -J/e$ between the charged current $j$ and the electric current $J$.

The explicit solutions for the charged component of the density $\delta n \equiv -\delta Q/e$, the neutral density $\delta\rho$, the electric field $E_y$, the electrostatic potential $\phi$, the neutral current $P_y$, and the electrically charged current $J_y$ are, respectively, as follows:

$$\delta n(y) = \frac{D_-}{(D_+^2 - D_-^2)(\varkappa_1^2 - \varkappa_2^2)}\frac{\sigma_-}{\sigma_+}\left(\varkappa_1\frac{\sinh\varkappa_1 y}{\cosh\frac{\varkappa_1 L}{2}} - \varkappa_2\frac{\sinh\varkappa_2 y}{\cosh\frac{\varkappa_2 L}{2}}\right)\frac{J}{e}\,, \tag{S90a}$$

$$\delta\rho(y) = -\frac{D_+}{(D_+^2 - D_-^2)(\varkappa_1^2 - \varkappa_2^2)}\frac{\sigma_-}{\sigma_+}\left[\left(\varkappa_1 - \frac{4\pi e\sigma_+}{\varkappa_1 D_+}\right)\frac{\sinh\varkappa_1 y}{\cosh\frac{\varkappa_1 L}{2}}\right.$$
$$\left.- \left(\varkappa_2 - \frac{4\pi e\sigma_+}{\varkappa_2 D_+}\right)\frac{\sinh\varkappa_2 y}{\cosh\frac{\varkappa_2 L}{2}}\right]\frac{J}{e}\,, \tag{S90b}$$

$$E_y(y) = \left[ \frac{1}{\sigma_+} - \frac{4\pi e D_-}{(D_+^2 - D_-^2)(\varkappa_1^2 - \varkappa_2^2)} \frac{\sigma_-}{\sigma_+} \left( \frac{\cosh \varkappa_1 y}{\cosh \frac{\varkappa_1 L}{2}} - \frac{\cosh \varkappa_2 y}{\cosh \frac{\varkappa_2 L}{2}} \right) \right] \frac{J}{e}, \tag{S90c}$$

$$\phi(y) = - \left[ \frac{y}{\sigma_+} - \frac{4\pi e D_-}{(D_+^2 - D_-^2)(\varkappa_1^2 - \varkappa_2^2)} \frac{\sigma_-}{\sigma_+} \left( \frac{1}{\varkappa_1} \frac{\sinh \varkappa_1 y}{\cosh \frac{\varkappa_1 L}{2}} - \frac{1}{\varkappa_2} \frac{\sinh \varkappa_2 y}{\cosh \frac{\varkappa_2 L}{2}} \right) \right] \frac{J}{e}, \tag{S90d}$$

$$P_y(y) = \frac{\sigma_-}{\sigma_+} \left\{ \frac{1}{\varkappa_1^2 - \varkappa_2^2} \left[ \varkappa_1^2 \frac{\cosh \varkappa_1 y}{\cosh \frac{\varkappa_1 L}{2}} - \varkappa_2^2 \frac{\cosh \varkappa_2 y}{\cosh \frac{\varkappa_2 L}{2}} \right. \right.$$
$$\left. \left. - 4\pi e \frac{D_+ \sigma_+ - D_- \sigma_-}{D_+^2 - D_-^2} \left( \frac{\cosh \varkappa_1 y}{\cosh \frac{\varkappa_1 L}{2}} - \frac{\cosh \varkappa_2 y}{\cosh \frac{\varkappa_2 L}{2}} \right) \right] - 1 \right\} \frac{J}{e}, \tag{S90e}$$

$$J_y(y) = J \equiv -ej. \tag{S90f}$$

The electrostatic potential (S90d) is related to the electric field (S90c) via Eq. (S53), and is normalized as follows: $\phi(0) = 0$. The linear part of the electrostatic potential (S90d) is straightforwardly determined by the Ohmic drift of the current $J$ while the non-linear part is related to the charge accumulation (S90a). As there is no simple relationship between these two contributions, the importance of one or another contribution cannot be determined from general reasoning. However, as we will see below, in a two-dimensional system, the Ohmic part of the electrostatic potential is determined by a harmonic function, while a non-harmonic contribution gives the charge accumulation part.

It is also instructive to write separately the densities of particles and holes given by the linear combinations (S26) and (S27) of the first two equations in Eq. (S90):

$$\delta n_e(y) = \frac{1}{2(D_+^2 - D_-^2)(\varkappa_1^2 - \varkappa_2^2)} \frac{\sigma_-}{\sigma_+} \frac{J}{e} \tag{S91a}$$
$$\times \left[ 4\pi e \sigma_+ \left( \frac{1}{\varkappa_1} \frac{\sinh \varkappa_1 y}{\cosh \frac{\varkappa_1 L}{2}} - \frac{1}{\varkappa_2} \frac{\sinh \varkappa_2 y}{\cosh \frac{\varkappa_2 L}{2}} \right) - D_h \left( \varkappa_1 \frac{\sinh \varkappa_1 y}{\cosh \frac{\varkappa_1 L}{2}} - \varkappa_2 \frac{\sinh \varkappa_2 y}{\cosh \frac{\varkappa_2 L}{2}} \right) \right],$$

$$\delta n_h(y) = \frac{1}{2(D_+^2 - D_-^2)(\varkappa_1^2 - \varkappa_2^2)} \frac{\sigma_-}{\sigma_+} \frac{J}{e} \tag{S91b}$$
$$\times \left[ 4\pi e \sigma_+ \left( \frac{1}{\varkappa_1} \frac{\sinh \varkappa_1 y}{\cosh \frac{\varkappa_1 L}{2}} - \frac{1}{\varkappa_2} \frac{\sinh \varkappa_2 y}{\cosh \frac{\varkappa_2 L}{2}} \right) - D_e \left( \varkappa_1 \frac{\sinh \varkappa_1 y}{\cosh \frac{\varkappa_1 L}{2}} - \varkappa_2 \frac{\sinh \varkappa_2 y}{\cosh \frac{\varkappa_2 L}{2}} \right) \right].$$

The solutions (S90) possess specific exciting properties which are worth discussing before proceeding to the analysis of a more complicated two-dimensional case.

In our experiments, the longest length $\varkappa_2^{-1}$ associated with the variation of the electric charge density is approximately one order of magnitude smaller than the linear size $L$ of our samples. Due to the hierarchy of the inverse length scales (S47), we fix, to mimic the realistic case, the length scales as $\varkappa_2 L \simeq L/\lambda_R = 10$ and consider the dependence of the electric charge density (S90a) on the other length scale $\varkappa_1^{-1}$. The corresponding charge accumulation patterns are shown in Fig. S9 for three values of $\varkappa_1$ which include, for the sake of generality and visualization, certain academic cases as well.

At large values of the second length $\varkappa_2^{-1}$ (i.e., at small $\varkappa_2$), there are two regions of space characterized by positive and negative values of the accumulated charge. As $\varkappa_2$ increases, the system develops a more complicated pattern: the electric charge density changes its sign three times, thus exposing four, instead of two, different regions with alternating order of the accumulated charge density. This pattern is a natural feature of the overlapping exponential tails with different amplitudes and characteristic lengths.

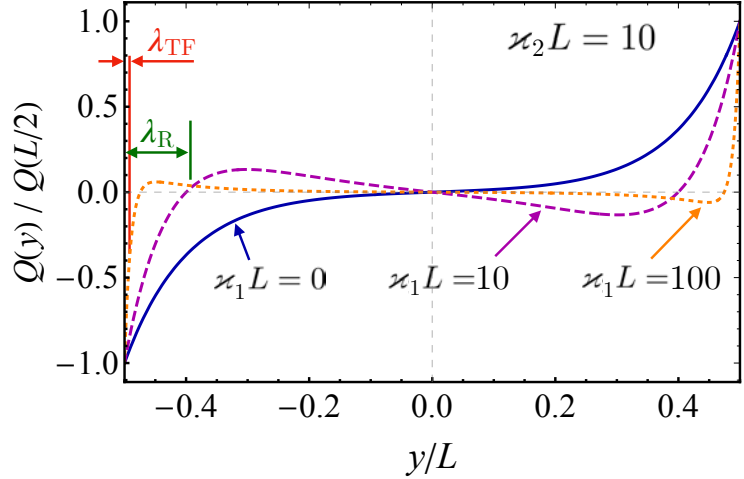

FIG. S9. Electric charge accumulation, $\delta Q = -e\delta n$, Eq. (S90a) at $L/\lambda_R \simeq \varkappa_2 L = 10$ and three values of the inverse length $\varkappa_1$. For $\varkappa_1 L = 100$ (the orange dotted line), the Thomas-Fermi length ($\lambda_{\mathrm{TF}} \simeq \varkappa_2^{-1} \equiv \lambda_2$) and the recombination length ($\lambda_R \simeq \varkappa_1^{-1} \equiv \lambda_1$) are shown explicitly. Experimentally realizable situation corresponds to large values of $\varkappa_1 L$ corresponding to four regions of charge accumulation (*cf.* also discusson around Figs. S10 and S11 below). The regions closest to the boundaries have width shorter than $\lambda_{\mathrm{TF}}$ and therefore they cannot be resolved in the experiment. The value $L/\lambda_R \simeq \varkappa_2 L = 10$ approximately corresponds to the conditions of our numerical experiment with $L \simeq 15\,\mu\mathrm{m}$ and $\lambda_R \simeq 1.4\,\mu\mathrm{m}$ along the $a$ axis of the WTe$_2$ crystal.

In Fig. S10 we show the appearance of the charge pattern as the function of two inverse length parameters $\varkappa_1$ and $\varkappa_2$. The two-region pattern appears at small values of $\varkappa_2$, and large values of $\varkappa_1$ while the charge pattern featuring the four regions is located at large values of both $\varkappa_1$ and $\varkappa_2$.

The alternating regions of charges along the $y$ direction of the sample are also illustrated in Fig. S11 for the fixed value of the second length with $\varkappa_2 L = 10$ as the function of the (inverse) first length, $\varkappa_1$. This plot features a $\Psi$-type figure which shows a two-region charge pattern at low $\varkappa_1$ and a four-region pattern at physical, higher values of $\varkappa_1$.

In two-component (semi)metals, the recombination length $\lambda_R$ is of the order of a micrometer while $\lambda_{\mathrm{TF}}$ amount a few nanometers. These lengths define the sign-flipping points of the charge accumulation regions as illustrated in Fig. S9. The physical point, shown by the red point in Fig. S10, belongs to the four-region phase for our (approximately) ten-micron-sized samples with very short Thomas-Fermi length $\lambda_{\mathrm{TF}} \simeq 1\,\mathrm{nm}$.

Our optical experiment cannot resolve the region in the nanometer scale so that the charge accumulation very close to the boundary avoids the experimental detection. Therefore, we detect the two-zone accumulation pattern rather than the four-zone one.

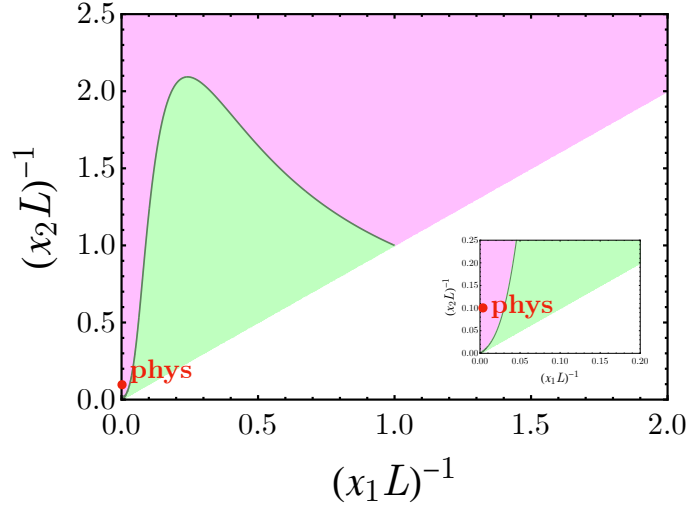

FIG. S10. The pattern of the charge accumulation in the $(\varkappa_1, \varkappa_2)$ plane of parameters. The regions where two (four) charge-alternating regions are realized correspond to green (magenta) colors. We show the region $\varkappa_1 > \varkappa_2$ corresponding the hierarchy (S46) and (S47). The red point approximately marks the physical case realized in the experiment: $(\varkappa_1 L)^{-1} \simeq \lambda_{\mathrm{TF}}/L \ll (\varkappa_2 L)^{-1} \simeq \lambda_R/L \simeq 0.1$, Eq. (S50). The inset shows the zooming in on the origin.

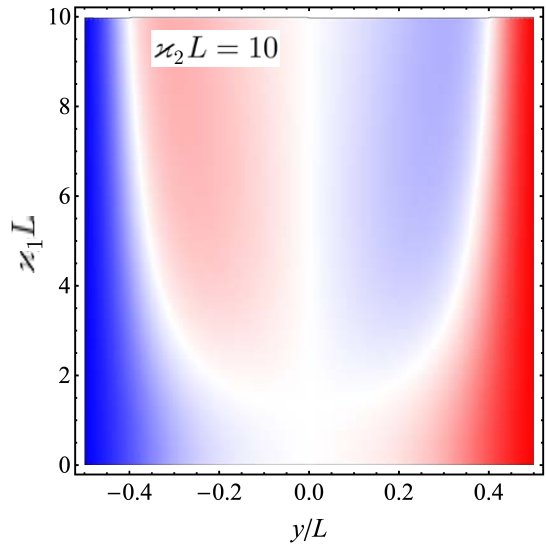

FIG. S11. The $\Psi$-type pattern of the charge accumulation along the conductor with the coordinate $y$ vs. the varying parameter $\varkappa_1$ at the fixed value of the combination length $\lambda_R/L \simeq \varkappa_2 L = 10$. The red (blue) color corresponds to the positively (negatively) charged regions. The physical case corresponds to a very large value of $\varkappa_1 L \simeq L/\lambda_{\mathrm{TF}} \gg 1$ when the near-surface charge accumulation region becomes undetectably thin and the experimentally observable accumulation pattern is given by two central regions.

Notice that the sign of the accumulated charge density depends not only on the direction of the electric current $J$ but also on the relative magnitude of the diffusion constants $D_e$ and $D_h$: For the same current $J$, the change in the sign of the difference $D_- = (D_e - D_h)/2$ also reverses the charge accumulation pattern to the opposite one. This property will be discussed later in detail.

### B. Charge accumulation: electron/hole asymmetry in conductivity and diffusivity

We have just seen that, in the general case, a passing electric current leads to the accumulation of electric charges at both sides of the sample. In this short subsection, we stress that the asymmetry in the conductivities of holes and electrons plays a decisive role in charge accumulation.

If the electron and hole conductivities are equal to each other, $\sigma_- = \sigma_e - \sigma_h = 0$, then, according to Eq. (S26), the electric current does not lead to any accumulation of charges and holes ($\delta n_e = \delta n_h = 0$) inside the sample. The quasiparticle current is absent as well:

$$\delta n = 0, \qquad \delta \rho = 0, \qquad P_y = 0 \qquad (\text{if } \sigma_e = \sigma_h). \tag{S92}$$

In this case, the electric field becomes a featureless coordinate-independent expression given by the standard Ohm law: $E = J/(e\sigma_+)$.

Equation (S90a) also indicates that the charge accumulation occurs if and only if the diffusion coefficients for electrons and holes are not equal to each other: $D_e \neq D_h$ so that $D_- \neq 0$. In the case when the diffusivities of the electrons and holes are the same $D_e = D_h$, while their conductivities are different, $\sigma_e \neq \sigma_h$, the electric charge accumulation is absent, $\delta n = 0$ and the electric field (S90c) is constant inside the sample. However, the accumulation of the neutral particles does occur (S90b): the sides of the sample will host a balanced neutral density of electrons and holes of equal magnitudes. The accumulation is maintained by the nonvanishing neutral particle current (S90e). We will consider this case in detail below.

The sign of the charge accumulation is determined by the direction of the current $J$, as well as by the relative magnitude of the conductivities of charges and holes. For example, the same current $J$ can induce a positive electric charge (with a dominant presence of holes) or a negative electric charge (with a dominant presence of electrons) at the vicinity of the same electrode depending on which conductivity, of charges or holes, is higher.

These observations can be interpreted as follows. The background electrostatic field $E_y$ generates the oppositely directed currents of electrons ($j_{e,y}$) and holes ($j_{h,y}$). If the conductivities of electrons and holes are not equal, $\sigma_e \neq \sigma_h$, then the magnitudes of these currents do not match each other. Therefore, a nonvanishing total particle current, $P_y = j_{e,y} + j_{h,y} \neq 0$, is generated (S90e). Since the particle current vanishes at the boundary of the sample, the nonvanishing bulk current produces a nonzero total particle density in the vicinity of the boundaries of the sample (S90b). This particle density is an electrically neutral quantity, implying that the neutral particle current brings the electrons and holes to the boundaries in the same proportions. However, if the diffusion rate of particles and holes are different, $D_e \neq D_h$, then these excitations diffuse differently from the boundaries to the bulk of the sample, thus leaving the corresponding charged imprint in the form of the nonzero charge density (S90a) in the vicinity of the boundary. It is the charge accumulation effect that we detect in the experiment. Thus, for the effect of the charge accumulation to be realized at the boundary, we need an electron-hole asymmetry both in their conductivities, $\sigma_e \neq \sigma_h$, and in their diffusivities, $D_e \neq D_h$.

Below, we set $\sigma_- > 0$, which corresponds to the case when the electron conductivity is higher than the conductivity of holes. In the case, $\sigma_- < 0$, the charge accumulation pattern remains precisely the same in its amplitude but gets reversed in its sign.

### C. Screening lengths and sign of the charge accumulation

Given by the theoretical expression (S90a), the charge accumulation could provide us with direct access to the evaluation of at least the longest one of the inverse lengths $\varkappa_1$ and $\varkappa_2$ from the experiment. As we will see below, our experimental data indicate that the slope of the charge accumulation near the boundaries can be excellently described by a single exponent, thus indicating that one of the lengths is much larger than another. This conclusion fits well our expectation (S50) that the parameters $1/\varkappa_1$ and $1/\varkappa_2$ correspond to the Thomas-Fermi and recombination lengths, respectively, with the hierarchy $\varkappa_1 \gg \varkappa_2$ which is expected to hold in two-component systems [13, 14].

Therefore, our experimental results justify our choice to consider the limit of large $\varkappa_1$ which simplifies the subsequent analysis. According to Eq. (S40), the spatial features of the solution are controlled by a single (inverse) length scale $\varkappa \equiv \varkappa_2 \simeq 1/\lambda_R$. Neglecting small near-boundary terms, we simplify the solution (S90) in the limit $\varkappa_2 \simeq 1/\lambda_R \ll \varkappa_1 \simeq \lambda_{\mathrm{TF}}^{-1}$.[1]

$$\delta n(y) \simeq -\frac{D_- \lambda_{\mathrm{TF}}^2}{(D_+^2 - D_-^2)\lambda_R} \frac{\sigma_-}{\sigma_+} \frac{\sinh \frac{y}{\lambda_R}}{\cosh \frac{L}{2\lambda_R}} \frac{J}{e}, \tag{S93a}$$

$$\delta\rho(y) \simeq \frac{D_+ \lambda_{\mathrm{TF}}^2}{(D_+^2 - D_-^2)} \frac{\sigma_-}{\sigma_+} \left(\frac{1}{\lambda_R} - \frac{4\pi e \lambda_R \sigma_+}{D_+}\right) \frac{\sinh \frac{y}{\lambda_R}}{\cosh \frac{L}{2\lambda_R}} \frac{J}{e}, \tag{S93b}$$

$$E_y(y) \simeq \left[\frac{1}{\sigma_+} + \frac{4\pi e D_- \lambda_{\mathrm{TF}}^2}{(D_+^2 - D_-^2)} \frac{\sigma_-}{\sigma_+} \frac{\cosh \frac{y}{\lambda_R}}{\cosh \frac{L}{2\lambda_R}}\right] \frac{J}{e}, \tag{S93c}$$

$$\phi(y) \simeq -\left[\frac{y}{\sigma_+} + \frac{4\pi e D_- \lambda_R \lambda_{\mathrm{TF}}^2}{(D_+^2 - D_-^2)} \frac{\sigma_-}{\sigma_+} \frac{\sinh \frac{y}{\lambda_R}}{\cosh \frac{L}{2\lambda_R}}\right] \frac{J}{e}, \tag{S93d}$$

$$P_y(y) \simeq -\frac{\sigma_-}{\sigma_+} \left\{\lambda_{\mathrm{TF}}^2 \left(\frac{1}{\lambda_R^2} - 4\pi e \frac{D_+\sigma_+ - D_-\sigma_-}{D_+^2 - D_-^2}\right) \frac{\cosh \frac{y}{\lambda_R}}{\cosh \frac{L}{2\lambda_R}} + 1\right\} \frac{J}{e}. \tag{S93e}$$

The expression of the electric current (S90f) remains, obviously, unchanged. Notice that the significant value of inverse length $\varkappa_2 \simeq 1/\lambda_{\text{TF}}$ leads to the diminishing of the amplitudes of the inhomogeneities in all discussed quantities, while the spatial size of the inhomogeneities is controlled by another, smaller quantity $\varkappa_1$ which remains finite.

---

[1] This equation is obviously valid for the bulk of the sample, at $|y| \leq L/2 - \lambda_{\mathrm{TF}}$, and not at the very narrow region of close to the boundary. Therefore, Eq. (S93e) is not valid at the points $y = \pm L/2$ that cannot be probed in our optical experiment anyway.

The densities of particles and holes (S91) become, respectively, as follows:

$$\delta n_e(y) = \rho_0 \left[ \frac{D_h}{\lambda_R} - \frac{4\pi e}{\varkappa_1}(\sigma_e + \sigma_h) \right] \frac{\sinh\frac{y}{\lambda_R}}{\cosh\frac{L}{2\lambda_R}}, \tag{S94a}$$

$$\delta n_h(y) = \rho_0 \left[ \frac{D_e}{\lambda_R} - \frac{4\pi e}{\varkappa_1}(\sigma_e + \sigma_h) \right] \frac{\sinh\frac{y}{\lambda_R}}{\cosh\frac{L}{2\lambda_R}}. \tag{S94b}$$

where the quantity

$$\rho_0 = \frac{\lambda_{\text{TF}}^2}{2D_e D_h} \frac{\sigma_e - \sigma_h}{\sigma_e + \sigma_h} \frac{J}{e}, \tag{S95}$$

characterizes the magnitude of the accumulated density of electrons and holes. We also used Eqs. (S29) and (S31).

Now let us simplify the solution further by considering the case when the diffusivities of the electrons and holes are the same, $D_e = D_h = D$ while their conductivities are different, $\sigma_e \neq \sigma_h$. In this limit, the symmetrized Thomas-Fermi length (S37) and the relaxation length (S38) become, respectively, as follows:

$$\varkappa^2 = \frac{1}{2}\left( \frac{1}{\lambda_+^2} + \frac{1}{\lambda_R^2} \right), \qquad \text{(for } D_e = D_h = D)$$

$$\lambda_+^2 = \frac{1}{4\pi e} \frac{D}{\sigma_e + \sigma_h} \equiv \lambda_{\text{TF}}^2, \qquad \lambda_R^2 = \sqrt{D\tau_R}. \tag{S96}$$

In agreement with our previous calculations, we find that for $D_e = D_h$ the electric charge accumulation is absent, $\delta n(y) = 0$, because the densities (S94) of electrons and holes coincide at each point of the sample, $\delta n_e(y) = \delta n_h(y)$. The electric field $E_y$ is a coordinate-independent quantity related to the electric current via the Ohm law: $J = e(\sigma_e + \sigma_h)E_y$. The quasiparticle density and the quasiparticle current (S93e) are, however, not vanishing.

The main conclusion arising from the analysis of Eq. (S93) is that in the background of the electric field, the two-carrier system accumulates the quasiparticle density (S93b) and generates the quasiparticle current (S93e) even in the absence of the electric charge accumulation (that is in the case when the conductivities of the electrons and holes are equal to each other $\sigma_e = \sigma_h$). Outside of the exact neutrality point, $\sigma_e \neq \sigma_h$, the electrically charged component of the system picks up the properties of its neutral component. While the quasiparticle density avoids, due to its neutrality, a direct detection by the electromagnetic probes, it is the electric charge density that allows us to find the indirect signatures of the neutral quasiparticle density and the neutral quasiparticle current. We use this property below to study the pseudo-hydrodynamic behavior of the neutral component of the electron-hole system.

According to the phase diagram of Fig. S10, the physical point $\varkappa_2 \simeq \lambda_{\text{TF}}^{-1} \gg \varkappa_1 \simeq \lambda_R^{-1}$, shown by the red dot in Fig. S11, corresponds to the accumulation of the electric charge close to the sides of the sample. In this limit, the additional sign flip of the electric charge in the multiple-alternating density pattern (the upper part of the diagram in Fig. S11) cannot be seen in the experiment due to the thinness of the boundary charge layer (which is of the order of $\lambda_{\text{TF}}$). We observe, instead, a single sign flip corresponding to the accumulation of the electric charge of the certain sign at one boundary and of the other sign at the opposite boundary.

We illustrate the normalized profiles of all the quantities (S93) for various relative diffusivities and conductivities in Fig. S12. The sign of the charge accumulation depends not only on the current direction but also on the relative magnitude of the diffusivities and conductivities of charges and holes. The electric current is carried by particles and holes traveling in opposite directions. When the electron and hole currents are imbalanced, the regions close to the contacts become predominantly saturated by a carrier of a single type, either by electrons or by holes. Due to the imbalance, the recombination of these excitations cannot remove all carriers and the electric charge of a single sign is therefore accumulated in these regions.

It is the relative magnitude of the electron-hole diffusivities, $D_e$ and $D_h$, and conductivities, $\sigma_e$, and $\sigma_h$, that determines the sign of the charge accumulation (S93a) near the boundaries at the fixed sign of the electric current. In other words, the accumulated charge (S93a) flips its sign if we pass from the higher diffusivity for electrons, $D_e > D_h$, to the higher diffusivity for holes, $D_e < D_h$ (with all other parameters fixed), while the Ohmic part of the potential stays the same.

Summarizing, in order for the charge accumulation (S90a) to occur, the electrons and holes must possess both non-equal diffusivities $D_e \neq D_h$ and non-equal conductivities, $\sigma_e \neq \sigma_h$. On the contrary, if any of these quantities are the same for particles and holes, the charge accumulations do not occur.

## S11. MISALIGNED CURRENT/CRYSTAL AXIS

In the previous Section, we demonstrated that in a two-carrier system with imbalanced electrons and holes, the background electric field leads to the longitudinal charge accumulation near the system's boundaries, even in the absence of a background magnetic field. The word "longitudinal" is used here to stress that the charge accumulation emerges in the direction of the electric field. While we studied the one-dimensional case, the effect also appears in the two-dimensional samples provided the electric field is aligned with one of the axes of the crystal while the boundary is perpendicular to the direction of the field as shown in Fig. S13a.

The longitudinal accumulation of electric charge is a close counterpart of the transverse charge accumulation in the Hall systems, which appears due to the emergence of the transverse voltage in the isotropic two-component conductors in the magnetic field background [13–15]. In our experiments, the transverse charge accumulation appears as well (see Figs. S8a and S8d) although the background magnetic field is absent while the Oersted magnetic field, generated by the electric current, is negligible. In this Section, we briefly argue that the transverse charge accumulation appears due to the intrinsic anisotropy of the crystal and the misalignment of the electric field axis concerning the boundaries.

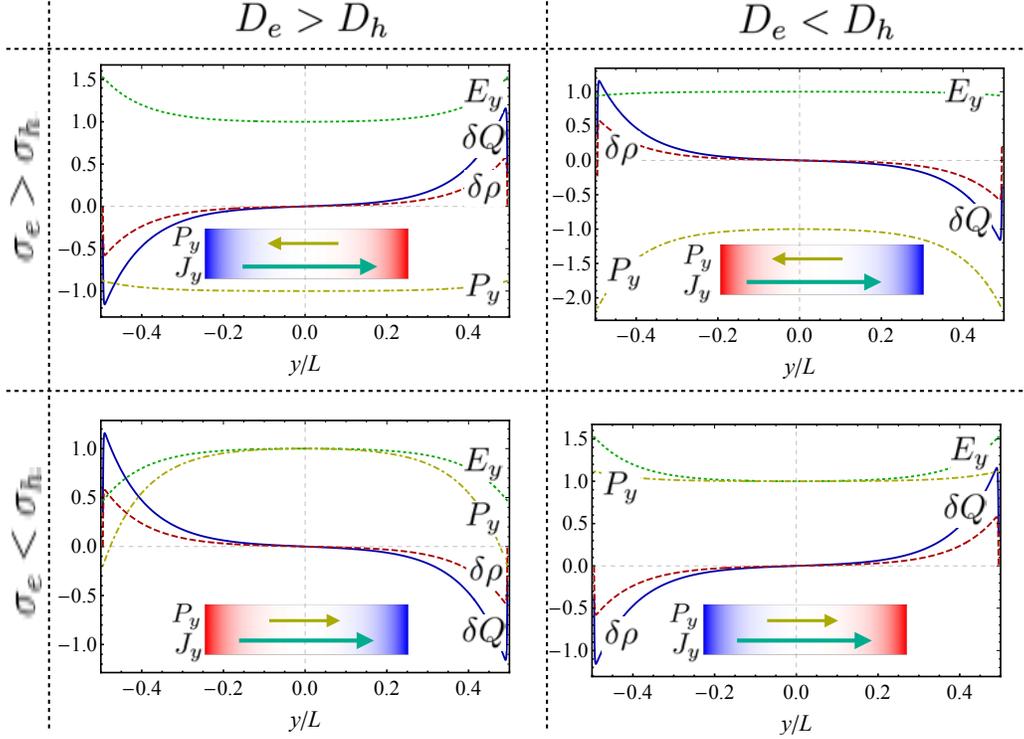

FIG. S12. Qualitative behavior of the electric charge density $\delta Q$, the neutral quasiparticle density $\delta\rho$, the quasiparticle current $P_y$ and the electric field $E_y$ in the sample with the coordinate $y$ along its length $L$. The left and right panels show the effect of the electron-hole diffusivity imbalance, $D_e > D_h$ and $D_e < D_h$, respectively. The upper and lower panels compare the effect of the imbalance in the electron-hole conductivities, with $\sigma_e > \sigma_h$ and $\sigma_e < \sigma_h$, respectively. The quantities $|P_y|$ and $|E_y|$ are normalized to unity at $y = 0$ while the densities of the neutral quasiparticle density $\delta\rho$ and the charged density $\delta n$ are shown in arbitrary units. The insets show the directions of electric and quasiparticle currents as well as the charge density pattern inside the sample (the experimentally unobservable part of the charge accumulation at the thin layers near the boundaries is not shown).

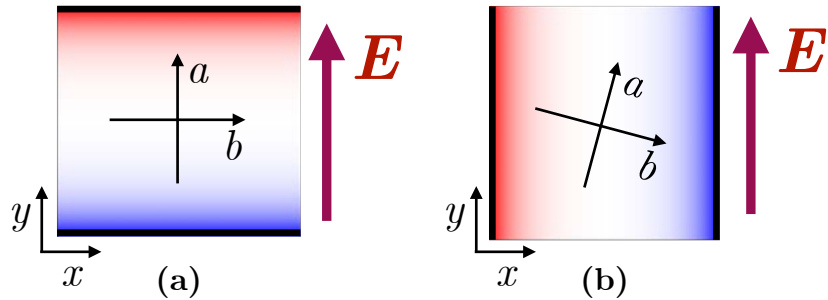

FIG. S13. (a) The longitudinal and (b) transverse charge accumulation near the boundaries (shown by the thick black lines) for (a) the perfectly oriented and (b) misaligned direction of the electric field $\boldsymbol{E}$ and the symmetry axis of the crystals in the $ab$ plane.

Consider the case of a spatially anisotropic crystal for which all quantities that characterize the dynamics of electrons and holes (conductivities, diffusion parameters, and, consequently, kinetic and recombination times, etc.) are spatially anisotropic. Then the transport equations (S35) get promoted to the matrix relations with the parameters $D_\pm$ and $\sigma_\pm$ that takes a matrix form. Finally, these relations are supplemented with the continuity equation for the electric current (S32), the recombination equation for the quasiparticle current (S33), and the Maxwell equation (S34) which are not affected by the anisotropy.

To simplify our considerations, let us assume that the anisotropy affects only the electric conductivity while leaving the diffusion parameter isotropic. In the basis of the orthogonal $a$ and $b$ axes, the electric conductivity matrix has the diagonal form

$$\hat{\sigma}^{(0)} = \begin{pmatrix} \sigma^{aa} & 0 \\ 0 & \sigma^{bb} \end{pmatrix}, \tag{S97}$$

where $\sigma^{aa}$ and $\sigma^{bb}$ are conductivities along the principal axes $a$ and $b$, respectively. The anisotropy implies that $\sigma^{aa} \neq \sigma^{bb}$.

Let us rotate the crystal counterclockwise around the $c$ axis by the angle $\theta$. The conductivity in the rotated sample is represented by the symmetric matrix,

$$\hat{\sigma} = \Omega^T \hat{\sigma}^{(0)} \Omega = \begin{pmatrix} \sigma^{xx} & \sigma^{xy} \\ \sigma^{yx} & \sigma^{yy} \end{pmatrix}, \tag{S98}$$

where the rotation in the $ab$ plane is given by the matrix

$$\Omega = \begin{pmatrix} \cos\theta & \sin\theta \\ -\sin\theta & \cos\theta \end{pmatrix}. \tag{S99}$$

The components of the conductivity matrix (S98) in the rotated crystal,

$$\sigma^{xx} = \sigma^{aa}\cos^2\theta + \sigma^{bb}\sin^2\theta, \tag{S100a}$$

$$\sigma^{yy} = \sigma^{aa}\sin^2\theta + \sigma^{bb}\cos^2\theta, \tag{S100b}$$

$$\sigma^{xy} = \sigma^{yx} = (\sigma^{aa} - \sigma^{bb})\cos\theta\sin\theta, \tag{S100c}$$

are, respectively, given by the longitudinal conductivities along the $x$ and $y$ axes and the off-diagonal conductivity.

Let take the electric field $\boldsymbol{E} = (0, E_y)$ parallel to the boundaries of the crystal in the $y$ direction as shown in Fig. S13b. For simplicity, let us assume an infinitely long slab in the $y$ direction and show that due to the misalignment of the crystals axes with the boundary and the electric field, the system accumulated electric charge in transverse directions.

It is clear that the currents and the densities of charges and quasiparticles, along with the electric field, are independent of the $y$ coordinate due to the translational symmetry in the longitudinal direction. Therefore, the electrostatic potential should be a linear function of the $y$ coordinate only, $\phi(x, y) = f_1 y + f_0$, where $f_0$ and $f_1$ are constants (if these constants were functions of the $x$ coordinate, then the electric field would depend on $y$ what contradicts the translational symmetry along $y$). Thus we arrive that the electric field along the $y$ axis, $E_y = -f_1$, does not depend on the transversal $x$ coordinate.

The electric charge conservation (S32) along with the independence of the electric current on the $y$ coordinate, $\boldsymbol{j}(x, y) = \boldsymbol{j}(y)$, implies the independence of the transverse current $j_x$ on the $x$ coordinate, $\partial_x j_x = 0$. However, the transverse current should be globally vanishing, $j_x = 0$, since the electric charge cannot quit the crystal at transverse boundaries. Then the $x$ component of Eq. (S35a), with vanishing right-hand side, leads to the relation between the $x$-independent parts of the electric fields in the transverse and longitudinal directions:

$$E_x^{(0)} = -\frac{\sigma_+^{xy}}{\sigma_+^{xx}} E_y \,. \tag{S101}$$

Here the superscript "(0)" reminds us that we consider only the $x$-independent component of the electric field. In the one-dimensional case considered earlier, this coordinate-independent component corresponds to the uniform electric field given by the first term in the square brackets of Eq. (S90c).

The transverse electric field (S101) produces – in a distant similarity to a Hall system [13, 14] and in close resemblance of the results on the one-dimensional case discussed in detail in the previous section – the charge accumulation in the transverse direction:

$$\delta n(x) = n_1 \sinh \kappa_1 x + n_2 \sinh \kappa_2 x \,, \tag{S102}$$

which shares similarity with its one-dimensional analogue (S90a). Here the inverse lengths $\kappa_1$ and $\kappa_2$ are given by Eqs. (S40) and (S46), where the conductivities are associated with the transverse diagonal components, $\sigma_\pm \equiv \sigma_\pm^{xx}$. To keep our considerations short, we do not pursue this discussion further, but we notice that the coefficients $n_1$ and $n_2$ that define the electric charge accumulation in Eq. (S102) can be calculated similarly to the approach described earlier in the one-dimensional case. It is interesting to notice that the behavior of the accumulated *electric charge* density is determined by the boundary conditions on the *neutral* quasiparticle currents. The charge accumulated patterns in two-dimensional geometries with finite-sized electrodes will be discussed below.

The essential message of this Section is that the coefficients $n_{1,2}$ are nonzero in the imbalanced system so that a misaligned crystal should exhibit the transverse charge accumulation (S102) illustrated in Fig. S13b. The same effect was also observed in our experiments in Figs. S8a and S8d which we attribute to the slight misalignment of the intrinsic crystal axis and the geometry of the sample. Below we consider a more involved system with finite-sized electrodes, whish leads us directly to the pseudo-hydrodynamic phenomena.

## S12. SOLUTION IN AN INFINITELY LONG SLAB

### A. General solution

We assume that the electric current is injected at the $y = -L/2$ (lower) side of the slab and then gets absorbed at the $y = L/2$ (upper) side as illustrated in Fig. S8. Following Ref. [16], it is instructive to consider the solution in a slab that is unbounded in the direction of the $x$ axis. In this case, it is convenient to represent all quantities, the densities and the currents, $\mathcal{O} = \mathcal{O}(\boldsymbol{r})$ in terms of the partial Fourier integral:

$$\mathcal{O}(x, y) = \int_{-\infty}^{\infty} \frac{dk}{2\pi} e^{ikx} \mathcal{O}_k(y), \tag{S103}$$

where $\mathcal{O}_k(y)$ are the Fourier coefficients which are given by the inverse Fourier transformation:

$$\mathcal{O}_k(y) = \int_{-\infty}^{\infty} dx\, e^{-ikx} \mathcal{O}(x, y), \tag{S104}$$

and $k \equiv k_x$ is the Fourier momentum along the $x$ axis.

Substituting into Eq. (S45) the representation (S103) for the charge density $\mathcal{O} = \delta n(\boldsymbol{r})$, we get the following equation for the Fourier coefficients:

$$\left(\partial_y^2 - \omega_{1,k}^2\right)\left(\partial_y^2 - \omega_{2,k}^2\right)\delta n_k(y) = 0, \tag{S105}$$

where we choose the positively-defined "frequencies":

$$\omega_a(k) = +\sqrt{\varkappa_a^2 + k^2}, \qquad a = 1, 2. \tag{S106}$$

A general solution of Eq. (S105) can be expressed in the following form:

$$\delta n_k(y) = \; n_{1,k}^+ e^{\omega_{1,k} y} + n_{1,k}^- e^{-\omega_{1,k} y} + n_{2,k}^+ e^{\omega_{2,k} y} + n_{2,k}^- e^{-\omega_{2,k} y} \equiv \sum_{a=1,2} \sum_{s=\pm} n_{a,k}^s e^{s\omega_{a,k} y}, \tag{S107}$$

where $n_{a,k}^{\pm}$ with $a = 1, 2$ are four arbitrary parameters which depend on the momentum $k$.

Using the set of equations (S36), one also gets the corresponding representation for the Fourier coefficient of the quasiparticle density:

$$\delta \rho_k(y) = \sum_{a=1,2} \sum_{s=\pm} \rho_{a,k}^s e^{s\omega_{a,k} y} \tag{S108}$$

where the parameters ($a = 1, 2$)

$$\rho_{a,k}^{\pm} = C_a^{(\rho)} n_{a,k}^{\pm}, \tag{S109}$$

with the coefficients $C_a^{(\rho)}$ given in Eq. (S83), are linked to the coefficients $n_{a,k}^{\pm}$ that determine the charge density (S107).

The electric field $\boldsymbol{E}$ satisfies the Maxwell equation (S34). Expressing the field $\boldsymbol{E}$ in terms of the electrostatic potential $\phi$ via Eq. (S53), the Maxwell equation becomes as follows:

$$\Delta\phi(\boldsymbol{r}) = 4\pi e \delta n(\boldsymbol{r}). \tag{S110}$$

The general solution of Eq. (S110) is given by the sum $\phi(\boldsymbol{r}) = \phi^{(n)}(\boldsymbol{r}) + \phi^{(\mathrm{h})}(\boldsymbol{r})$ of a solution $\phi_h$ of the inhomogeneous equation (S110) and an arbitrary harmonic function $\phi_h$ which solves the Laplace equation:

$$\Delta\phi^{(\mathrm{h})}(\boldsymbol{r}) = 0. \tag{S111}$$

The harmonic part of the solution can also be represented in the form (S103) with the coefficients:

$$\phi_k^{(\mathrm{h})}(y) = \Phi_k^+ e^{\omega_{0,k} y} + \Phi_k^- e^{-\omega_{0,k} y}, \tag{S112}$$

and

$$\omega_{0,k} = k\,. \tag{S113}$$

To keep our equations simple, we allow $\omega_{0,k}$ to take any sign in contrast to the previously defined frequencies (S106). As we will see below, the (non-)harmonic component of the solution in two spatial dimensions is the analog of the (non-)homogeneous part of the one-dimensional solution discussed earlier.

In the following, we represent $\phi_n$ in terms of the Fourier integral (S103). The explicit solution of the Maxwell equation (S110) with the source density (S107) reads as follows:

$$\phi_k(y) = \sum_{a=0}^{2}\sum_{s=\pm} \phi_{a,k}^s e^{s\omega_{a,k} y}, \tag{S114}$$

where the non-harmonic coefficients $\phi_{1,k}^\pm$ and $\phi_{2,k}^\pm$ are again related to the charge density (S107) via the following relation:

$$\phi_{a,k}^\pm = \frac{4\pi e}{\varkappa_a^2} n_{a,k}^\pm\,, \qquad (a = 1, 2)\,, \tag{S115}$$

while the harmonic coefficients $\phi_{0,k}^\pm \equiv \Phi_k^\pm$ of the electrostatic potential remain still undetermined at this stage.

The Fourier components $\boldsymbol{E}_k(y)$ of the electric field (S53) then become rigidly fixed:

$$E_{k,x}(y) = -ik\phi_k(y) \equiv -ik\sum_{a=0}^{2}\sum_{s=\pm}\phi_{a,k}^s e^{s\omega_{a,k} y}, \tag{S116a}$$

$$E_{k,y}(y) = -\partial_y\phi_k(y) \equiv -\sum_{a=0}^{2}\sum_{s=\pm} s\omega_{a,k}\phi_{a,k}^s e^{s\omega_{a,k} y}. \tag{S116b}$$

The electric current $\boldsymbol{j}$ and the particle current $\boldsymbol{P}$ can now be unambiguously determined from the set of equations (S35):

$$\boldsymbol{j} = -D_+\boldsymbol{\nabla}\delta n - D_-\boldsymbol{\nabla}\delta\rho - \sigma_+\boldsymbol{E}, \tag{S117a}$$

$$\boldsymbol{P} = -D_-\boldsymbol{\nabla}\delta n - D_+\boldsymbol{\nabla}\delta\rho - \sigma_-\boldsymbol{E}, \tag{S117b}$$

which involve all the quantities $\delta n$, $\delta\rho$ and $\boldsymbol{E}$ that were identified above. We remind that, according to Eq. (S27), the charged current $\boldsymbol{j}$ has the opposite direction with respect to the electric current $\boldsymbol{J} = -e\boldsymbol{j}$.

Substituting Eqs. (S107), (S108), and (S116) into Eq. (S117a), we notice that the charged current $\boldsymbol{j}$ is determined only by the harmonic component of the electric field:

$$\boldsymbol{j}(\boldsymbol{r}) = -\sigma_+\boldsymbol{E}^{(\mathrm{h})}(\boldsymbol{r}) \equiv \sigma_+\boldsymbol{\nabla}\phi^{(\mathrm{h})}(\boldsymbol{r})\,, \tag{S118}$$

with the following Fourier coefficients:

$$j_{x,k}(y) = ik\sigma_+\left(\Phi_k^+ e^{ky} + \Phi_k^- e^{-ky}\right)\,, \tag{S119a}$$

$$j_{y,k}(y) = k\sigma_+\left(\Phi_k^+ e^{ky} - \Phi_k^- e^{-ky}\right)\,. \tag{S119b}$$

This property qualitatively agrees with the one-dimensional case in which the charged current $\boldsymbol{j} \equiv (0, j_y)$ is determined only by a coordinate-independent (that is, harmonic, in one space dimension) component of the electric field (S78).

It is interesting to observe that while the charged current originates from the purely harmonic functions (S119), the charged density is entirely determined by the non-harmonic expressions (S107). These seemingly unrelated terms are both contributing to the quasiparticle current (S117b) which can be represented as a sum of the harmonic and non-harmonic components, respectively:

$$\boldsymbol{P} = \boldsymbol{P}^{(\mathrm{nh})} + \boldsymbol{P}^{(\mathrm{h})}\,. \tag{S120}$$

The harmonic component is the conserved part of the quasiparticle current which is proportional to the electric current:

$$\boldsymbol{P}^{(\mathrm{h})} = \frac{\sigma_-}{\sigma_+}\boldsymbol{j}\,. \tag{S121}$$

The system inevitably produces the Ohmic flow of quasiparticles due to the imbalance in the electron-hole conductivities ($\sigma_e \neq \sigma_h$). The Ohmic flow cannot generate the quasiparticle density and cannot lead to the electric charge accumulation in the system because the Ohmic (harmonic) part of the quasiparticle current is exactly conserved,

$$\boldsymbol{\nabla}\cdot\boldsymbol{P}^{(\mathrm{h})} \equiv 0\,. \tag{S122}$$

The non-harmonic part of the quasiparticle current (S120) is given by

$$P_{x,k}^{(\mathrm{nh})}(y) = -ik \sum_{a=1,2} \sum_{s=\pm} p_{a,k}^s e^{s\omega_{a,k}y}, \tag{S123a}$$

$$P_{y,k}^{(\mathrm{nh})}(y) = -\sum_{a=1,2} \sum_{s=\pm} s\omega_{a,k} p_{a,k}^s e^{s\omega_{a,k}y}, \tag{S123b}$$

where, similarly to the one-dimensional case, the new coefficients

$$p_{a,k}^{\pm} = C_a^{(P)} \, n_{a,k}^{\pm}, \qquad a = 1, 2, \tag{S124}$$

can be calculated with the use of Eqs. (S37), (S109), and (S115). The coefficients $C_a^{(P)}$ are given by Eq. (S86).

The non-harmonic part (S123) represents the non-Ohmic component of the neutral quasiparticle current (S120) which leads to the accumulation of the electric charge observed in the experiment. We will see below that the Ohmic and non-Ohmic components of the quasiparticle current are related to each other via the boundary conditions on the neutral quasiparticle current $\boldsymbol{P}$.

## B. Boundary conditions

Identifying the normally injected (or taken away) currents at the upper and lower sides of the strip

$$I^{\pm}\left(x, y = \pm\frac{L}{2}\right) = j_y^{\pm}\left(x, y = \pm\frac{L}{2}\right). \tag{S125}$$

and implementing Eq. (S104) to rewrite Eq. (S125) in the Fourier space,

$$I_k^{\pm} = \int_{-\infty}^{\infty} dx\, e^{-ikx} j_y\left(x, y = \pm\frac{L}{2}\right), \tag{S126}$$

we determine the coefficients $\phi_{0,k}^{\pm} \equiv \Phi_k^{\pm}$ via Eq. (S119b) using the relations:

$$I_k^{\pm} = j_{y,k}\left(x, y = \pm\frac{L}{2}\right) \equiv k\sigma_+ \left(\Phi_k^+ e^{\pm kL/2} - \Phi_k^- e^{\mp kL/2}\right). \tag{S127}$$

Inverting this matrix equation, we fix the harmonic coefficients of the solution as follows:

$$\Phi_k^{\pm} = \frac{1}{2\sigma_+ k \sinh kL} \left(e^{\pm kL/2} I_k^+ - e^{\mp kL/2} I_k^-\right). \tag{S128}$$

Notice that the solution is self-consistent if and only if the global conservation of electric charge is respected. Indeed, if the total injected charge and the total charge taken away are equal to each other,

$$\int\limits_{-\infty}^{\infty} dx\, j_y\left(x, y = +\frac{L}{2}\right) = \int\limits_{-\infty}^{\infty} dx\, j_y\left(x, y = -\frac{L}{2}\right), \tag{S129}$$

then

$$\lim_{k \to 0}\left(I_k^+ - I_k^-\right) = 0, \tag{S130}$$

or $I_k^+ = I_k^- + O(k^1)$ as $k \to 0$. We get for the parameters $\phi_{0,k}^\pm = O(k^{-1})$ so that the Fourier coefficients (S119) are finite in the infrared limit, $j_{y,k}(y) = O(k^0)$, implying that the electric currents $\boldsymbol{j}(\boldsymbol{r})$ are not divergent inside the slab due to the infinite charge accumulation caused by an imbalanced source and drain flows of the electric current.

The electric current (S118) is given by the harmonic components (S119) which are completely determined – in our setup – via the boundary conditions for the electric current (S128). On the other hand, the electric charge accumulation and quasiparticle distributions possess also a non-harmonic component which remains undetermined at this stage. More concretely, we are now left with the task to find the four remaining coefficients $n_{1,2}^\pm$ which determine the charge density (S107). These coefficients can subsequently be used to express the coefficients for the neutral quasiparticle density $\rho_{1,2}^\pm$, the electric potential $\phi_{1,2}^\pm$, and the neutral quasiparticle current $p_{1,2}^\pm$ via the relations (S109), (S115), and (S86), respectively.

While the intrinsically harmonic electric current is entirely defined by its Ohmic flow, the non-harmonic components of the quasiparticle current can only be constrained by the behavior of the quasiparticle current at the boundaries. The quasiparticle current $\boldsymbol{P}$ possesses both harmonic and non-harmonic components, with the harmonic component linking tightly to the electric current (S121). Therefore, the non-harmonic part of $\boldsymbol{P}$ can only be fixed by certain boundary constraints.

The first pair of equations for the non-harmonic coefficients $n_{1,2}^\pm$ is set by the boundary condition that the quasiparticle current $\boldsymbol{P}$ cannot quit the sample at the upper and lower edges of the slab:

$$P_y\left(x, y = \pm\frac{L}{2}\right) = 0. \tag{S131}$$

Written in the Fourier space, the condition (S131) for the normal component reads as follows: $P_{y,k}(y = \pm L/2) = 0$.

The remaining two relations on $n_{1,2}^\pm$ should, in principle, be set by the conditions to the tangential $P_x$ component of the quasiparticle current at the upper and lower edges. What are these conditions? In our idealized setup, the harmonic part of $\boldsymbol{P}$ cannot be constrained at all since it is tightly linked, via the system of Eqs. (S35), to the Ohmic flow of the electric current. Moreover, similarly to the electric current, it is easy to check that the tangential component of the harmonic part of the quasiparticle current $\boldsymbol{P}$ is not vanishing at the boundary similarity to the tangential component of the charged current. This property becomes evident if we notice that the harmonic parts of both currents constitute the same Ohmic flow.

It is not clear *ab initio* what should be the constraints on the tangential non-harmonic component of the quasiparticle current at the boundary. Restricting ourselves to the most straightforward cases, one could assume that this current component is vanishing at the border or equals the minus harmonic component (in the latter case, the total tangential current vanishes at the boundary). One can also generalize this condition by assuming that the tangential parts of both components, given in the Fourier images by Eqs. (S120), are proportional to each other:

$$\left[ P_x^{(\text{nh})} + g P_x^{(\text{h})} \right]\Big|_{y=\pm L/2} = 0 \,. \tag{S132}$$

The coefficient of proportionality, $g$, determines phenomenologically the property of the boundary that (de)couples the harmonic (h) and non-harmonic (nh) components. The two simplest cases, mentioned above, correspond to $g=0$ and $g=1$, respectively.

In general case, the set of the boundary equations for the tangential component of the quasiparticle current becomes as follows (with $s, r = \pm 1$):

$$\sum_{a=1}^{2} \sum_{s=\pm} p_{a,k}^s e^{sr\omega_{a,k} L/2} = g\sigma_- F_{+,k}^r \,, \tag{S133a}$$

$$\sum_{a=1}^{2} \sum_{s=\pm} s\omega_{a,k} p_{a,k}^s e^{sr\omega_{a,k} L/2} = \sigma_- k F_{-,k}^r \,, \tag{S133b}$$

where the following combinations give the functions on the right-hand side:

$$F_{s,k}^r = \Phi_k^+ e^{rkL/2} + s\Phi_k^- e^{-rkL/2}, \qquad s, r = \pm 1 \,. \tag{S134}$$

Equations (S133a) and (S133b) set the boundary conditions for the tangential and normal components of the quasiparticle current.

The explicit form of the boundary conditions (S133) for the quasiparticle current is as follows:

$$p_{1,k}^+ e^{\omega_{1,k} L/2} + p_{1,k}^- e^{-\omega_{1,k} L/2} + p_{2,k}^+ e^{\omega_{2,k} L/2} + p_{2,k}^- e^{-\omega_{2,k} L/2} = g\sigma_- F_+^+, \tag{S135a}$$

$$p_{1,k}^+ e^{-\omega_{1,k} L/2} + p_{1,k}^- e^{\omega_{1,k} L/2} + p_{2,k}^+ e^{-\omega_{2,k} L/2} + p_{2,k}^- e^{+\omega_{2,k} L/2} = g\sigma_- F_+^-, \tag{S135b}$$

$$p_{1,k}^+ \omega_{1,k} e^{\omega_{1,k} L/2} - p_{1,k}^- \omega_{1,k} e^{-\omega_{1,k} L/2} + p_{2,k}^+ \omega_{2,k} e^{\omega_{2,k} L/2} - p_{2,k}^- \omega_{2,k} e^{-\omega_{2,k} L/2} = k\sigma_- F_-^+, \tag{S135c}$$

$$p_{1,k}^+ \omega_{1,k} e^{-\omega_{1,k} L/2} - p_{1,k}^- \omega_{1,k} e^{\omega_{1,k} L/2} + p_{2,k}^+ \omega_{2,k} e^{-\omega_{2,k} L/2} - p_{2,k}^- \omega_{2,k} e^{\omega_{2,k} L/2} = k\sigma_- F_-^-. \tag{S135d}$$

The right-hand side of this matrix equation is related to the harmonic modes of the solution and is fixed by the form of the injected and absorbed electric current (S128). The coefficients $\Phi_k^\pm$ at this side also determine the lines of the electric current (S119) inside the sample.

The left-hand determines of Eq. (S135) gives the coefficients $p_{a,k}^\pm$ which, along with the coefficients $\Phi_k^\pm$, provide us with the density of the electric charge (S107), the neutral quasiparticle density (S108), the lines of the electric fields (S116), and the neutral quasiparticle currents (S135) via the relations (S109), (S115), and (S86). Thus, we get the complete description of the two-component transport system.

### C. Pointlike sources

Let us consider the $\delta$-function sources located, respectively, at the points $x = +a/2$ and $x = -a/2$ of the upper, $y = +L/2$, and lower, $y = -L/2$, edges of the slab:

$$j_y\left(x, y = \pm \frac{L}{2}\right) = I_0 \delta\left(x \mp a/2\right) . \tag{S136}$$

The Fourier components of the normal boundary currents (S126),

$$I_k^\pm(a) = I_0 e^{\mp ika/2} , \tag{S137}$$

give us the following harmonic coefficients of the solution (S128):

$$\Phi_k^\pm = \frac{I_0}{\sigma_+ k \sinh kL} \sinh \frac{\pm kL - ika}{2}, \tag{S138}$$

thus leading us to the following source functions (S134):

$$F_{+,k}^+(a) = \frac{I_0}{\sigma_+ k} \frac{e^{-ika/2} \cosh kL - e^{ika/2}}{\sinh kL}, \tag{S139a}$$

$$F_{+,k}^-(a) = \frac{I_0}{\sigma_+ k} \frac{e^{-ika/2} - e^{ika/2} \cosh kL}{\sinh kL}, \tag{S139b}$$

$$F_{-,k}^\pm(a) = \frac{I_0}{\sigma_+ k} e^{\mp ika/2} . \tag{S139c}$$

At the head-on location of the point-like sources, $a = 0$, the harmonic coefficients (S128) and the source functions (S134) are respectively, as follows:

$$\left.\Phi_k^\pm(a)\right|_{a=0} = \pm \frac{I_0}{2k\sigma_+ \cosh kL/2}, \tag{S140}$$

$$\left.F_{+,k}^\pm a\right|_{a=0} = \pm \frac{I_0}{\sigma_+ k} \tanh \frac{kL}{2}, \tag{S141}$$

$$\left.F_{-,k}^\pm a\right|_{a=0} = \frac{I_0}{\sigma_+ k} . \tag{S142}$$

The electric current (S118) is determined by the harmonic component of the electrostatic potential which takes the following form in the mixed coordinate-momentum representation:

$$\phi_k^{(\mathrm{h})}(y) = \frac{I_0 \sinh(ky)}{k\sigma_+ \cosh(kL/2)} \,. \tag{S143}$$

The solution in the coordinate space can be obtained by application of the Fourier transform (S103) to the above expression.

### D. Finite-sized electrodes

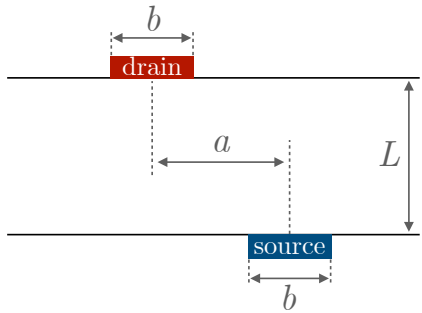

FIG. S14. Geometry of electrodes and samples.

Consider now the extended electrodes of the width $b$ each, centered at the points $+a/2$ and $-a/2$ at upper (lower) edges of the band, Fig. S14. For reference, the experimentally observed charge density maps in Fig. 2c of the main text are produced with the vanishing the offset $a = 0$ between the electrodes of the width $b = 2.5\,\mu\mathrm{m}$ separated by the distance $L = 15\,\mu\mathrm{m}$, while the density maps of Fig. 2f can be reproduced in the same geometry but with the offset about $a = 0.2\,\mu\mathrm{m}$.

The electric current at the edges is then:

$$j_y\left(x, y = \pm\frac{L}{2}\right) = \begin{cases} I_0/b, & \left|x \mp \frac{a}{2}\right| \leqslant \frac{b}{2}\,, \\ 0, & \text{otherwise}\,, \end{cases} \tag{S144}$$

so that the total current entering (leaving) the sample at lower (upper) edges is $I_0$. The Fourier transform of the current with the finite-sized electrodes (S144) is

$$I_k^{\pm}(a, b) = I_0 e^{\mp ika/2} \cdot \frac{2}{bk} \sin\frac{bk}{2}\,, \tag{S145}$$

which differs from the point-like source (S137) by multiplication by the factor $2\sin(bk/2)/(bk)$. Consequently, the functions $F_{\pm,k}^{\pm}$ and $\Phi_k^{\pm}$ for the finite-sized extended electrodes (S144) are easily obtained from the expressions for $\Phi_k^{\pm}(a)$ given in Eq. (S138) and $F_{\pm,k}^{\pm}(a)$, Eq. (S139), that correspond to the point-like electrodes:

$$\Phi_k^{\pm}(a, b) = \frac{2}{bk} \sin \frac{bk}{2} \Phi_k^{\pm}(a), \tag{S146}$$

$$F_{\pm,k}^{\pm}(a, b) = \frac{2}{bk} \sin \frac{bk}{2} F_{\pm,k}^{\pm}(a). \tag{S147}$$

The electric current is determined solely by the harmonic component which takes, for the wide electrodes of the finite width $b$ and zero offset $a = 0$, the following form:

$$\phi_k^{(h)}(y) = \frac{2I_0}{b\sigma_+ k^2} \frac{\sinh(ky)\sin(bk/2)}{\cosh(kL/2)}. \tag{S148}$$

## S13. DENSITIES, CURRENTS, AND VORTICES

### A. Overall picture

In this Appendix, we discuss an electron-hole system with specially chosen parameters that were presented, at one hand, to enhance the geometric effects and make the presence of the backflow in the quasiparticle flow visually appealing, and, at the other hand, to match the experimental setup closely. Unfortunately, given the multitude of different parameters which describe the two-dimensional solution (involving, notably, unknown conductivities $\sigma_e$, $\sigma_h$, the diffusion coefficients $D_e$, $D_h$ and the boundary parameter $g$), the accuracy of our experiment does not allow us to fix them reliably apart from the scale of the recombination length around $\lambda_R \sim 1\,\mu$m within the range $\lambda_R \simeq (0.5 - 2.0)\,\mu$m. Moreover, the theoretical solutions cannot be obtained in a closed analytical form, making the analysis even more complicated. However, we found via matching of the results of the numerical analysis with the experimental data that in a particular wide scale of these parameters, the theoretical result for the geometry of the charge accumulation region matches, albeit inevitably significant systematical uncertainty, the experiment very well.

We consider an infinitely long strip of the width $L$ with two finite-width electrodes, the source $(+I_0)$ and the sink $(-I_0)$, placed in front of each other at, respectively, the lower and upper edges of the strip. In notations of Fig. S14, we choose the offset $a = 0$ and the width of electrodes $b = 0.16L \simeq 2.5\,\mu$m at $L \simeq 15\,\mu$m. We choose the lengths scales $\varkappa_1 L = 300$ and $\varkappa_2 L = 10$ implying $\varkappa_1^{-1} \simeq 50\,$nm and $\varkappa_2^{-1} \simeq 1.5\,\mu$m. According to our identification (S50), the second length scale is close to the recombination length, $\varkappa_2^{-1} \simeq \lambda_R$ along the $a$ crystal axis in an approximate agreement with the value obtained in our experiment, Fig. S5. In principle, the first length scale, $\varkappa_1^{-1}$, should be of the order of the Thomas-Fermi length which is about $1\,$nm. However, the exact value of the first length scale is not important for the qualitative long-scale features of our solution as long as it is much shorter than the recombination length, $\varkappa_1^{-1} \ll \varkappa_2^{-1}$. Therefore, for the sake of the numerical convenience, we choose the larger value of $\varkappa_1^{-1}$.

We set the boundary parameter for the quasiparticle current to a small but nonzero value,

$g = 0.01$, which implies, according to Eq. (S132) that at the boundary, the non-harmonic component of the quasiparticle current equals to a small fraction (fixed to be just 1%) of the harmonic component. As we discussed earlier, the phenomenological parameter $g$ is a material property that cannot be fixed ab initio. The presence of the transverse boundary does not affect the results noticeably as long as the width of the sample is much larger than the relaxation length $\lambda_R$ (in our case, the ratio is about 20).

In order to support the electric charge accumulation in the vicinity of the electrodes, we require – following the one-dimensional example that we considered earlier – that the electrons and hole possess different diffusion consants and different conductivities: $D_e \neq D_h$ and $\sigma_e \neq \sigma_h$. The solution is determined with the help of the boundary conditions (S135) that fix the parameters $p_{1,2}^\pm$ of the quasiparticle flow. The latter coefficients are related to the electric charge components $n_a^\pm$ via the relation (S124) and the coefficients $C_{1,2}^{(P)}$ given in Eq. (S86) as an involved function of the diffusion coefficients and conductivities for electrons and holes. We choose for the ratio $C_2^{(P)}/C_1^{(P)} = 10^2$ which matches the hierarchy $\varkappa_1^{-1} \sim \lambda_{\mathrm{FT}} \ll \lambda_R \sim \varkappa_2^{-1}$. The influence of the length-scale hierarchy, $\varkappa_1^{-1}$ and $\varkappa_2^{-1}$, as well as the value of the boundary condition parameter $g$ on the geometry of the charge accumulation domains and the quasipaticle currents will be briefly discussed closer to the end of this section.

This set of parameters allows us to model the geometrical picture of the electric charge accumulation, which correctly matches the optical experiment. The comparison of experimental results and the theoretical model is shown in Fig. 2 of the main text. In this Supplementary material, we discuss other characteristics of the system which accompany the charge accumulation. As an example, we consider the case of vertically positioned electrodes shown in Figs. 2a,b,c and depict in Fig. S15 the whole variety of charge and quasiparticle densities and currents in the bulk of the sample.

Figure S15a represents the electric charge density which appears already in Fig. 2c of the main text. It can be compared with the corresponding quasiparticle density, which is shown in Fig. S15b. Both charged and neutral densities exhibit qualitatively similar alternating-sign patterns around the electrodes, albeit slightly different geometrical structures. The streamlines of the electric current, shown in Fig. S15c, are entirely given by the featureless Ohmic drift, which is governed by the harmonic component of the electrostatic potential. For completeness, the colors in the same figure show the electrostatic potential.

The quasiparticle current, Fig. S15d (Fig. 3a of the main text), however, exhibits a nontrivial behavior featuring a pair of the whirlpool-like structures in the bulk of the sample and a backflow near the electrodes. While these whirlpool-like structures are distantly similar to the hydrodynamic vortices in fluids, the analogy is not complete because the ordinary hydrodynamic flow is associated with the conserved flow while the quasiparticle charge is not conserved. As a result, the quasiparticle current streamlines are not closed: while a part of the quasiparticle flow originates and terminates at the source and the drain, some quasiparticle current lines terminate at the edge, leading to the charge accumulation as it is described in the main text. A similar effect leads to a later electric charge accumulation in almost-compensated conductors in the Hall geometry in the presence of the background magnetic field [13–15]. Figure S15 allows us to visualize different features of the alternating-sign charge accumulation mechanism due to the "pseudo-hydrodynamic" quasiparticle flow in the two-dimensional geometry.

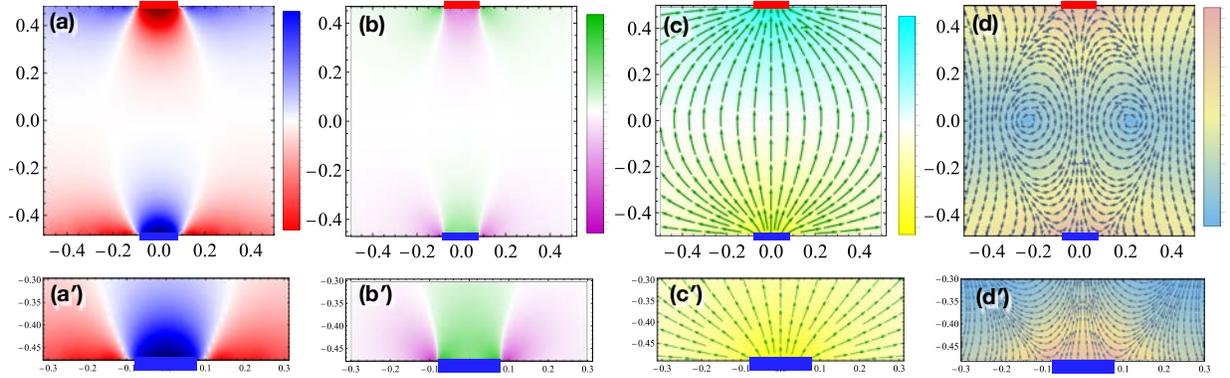

FIG. S15. Charged and quasiparticle densities and currents in the two-component conductor with the wide electrodes attached at the top $(x, y) = (0, +L/2)$ and the bottom $(x, y) = (0, -L/2)$ edges: (a) the electric charge density, (b) the quasiparticle density, (c) the streamlines of electric charge current, and (d) the streamlines of the non-Ohmic component of the quasiparticle current. The colors in (a), (b) and (c), (d) show, respectively, the magnitude of the densities and currents. The streamlines of the Ohmic component of the quasiparticle current are identical to those of the charged current shown in panel (c). All lengths are given in units of the width of the strip $L$ with the parameters discussed in the text. The magnitudes of all quantities are presented in arbitrary units. The figures a′, …, d′ zoom in the region in the close vicinity of the bottom electrode.

The difference in the electrostatic potentials between the source and the drain generates the harmonic Ohmic flow of the electric charge, Fig. S15c. The charged harmonic flow does not lead directly to the charge accumulation. However, in a close analogy with the one-dimensional example discussed earlier, the charged current produces the collinear quasiparticle current across the slab. In the vicinity of the electrodes, the non-Ohmic part of the quasiparticle flow creates the backflow close to the edges of the sample, Fig. S15d. The quasiparticle backflow, directed against the main flow, brings the quasiparticles back to the source (drain) located at the edge of the sample. The latter effect leads to the accumulation of the net quasiparticle number at the edges, Fig. S15b. The Ohmic part of the quasiparticle current (not shown) represents a conserved (non-compressible) flow that does not lead to density accumulation.

The carriers, accumulated at the boundaries, tend to spread back towards the bulk of the sample due to the diffusion. While the quasiparticle current brings equal electrons and holes towards the boundary, these charge carriers diffuse differently due to the imbalanced diffusion constants. The latter effect produces the alternating charge and particle densities in the vicinity of the electrodes, shown in Fig. S15a and Fig. S15b, respectively.

Since the quasiparticle number is not a conserved quantity, the harmonicity condition does not apply to the quasiparticle current. This property explains, in particular, the qualitative difference between the charged flow in Fig. S15c and the quasiparticle current in Fig. S15d. Notice that that normal component of both charged and quasiparticle currents vanishes at the boundary: while the streamlines of the quasiparticle current in Fig. S15d approach the edges normally, the magnitude of its normal component vanishes at the edge.

Remarkably, the quasiparticle backflow near the electrodes is also associated with the appearance of the symmetric pair of the whirlpools of the neutral quasiparticle current in the bulk of the sample, Fig. S15d. In the center of each of these whirlpools, the quasiparticle flow vanishes. The centers are located in the region where the electric charge density, Fig. S15a, and the neutral quasiparticle charge density, Fig. S15b, are vanishing as well. On the other hand, the whirlpool structures of the quasiparticle flow of the non-Ohmic (non-harmonic) component of the quasiparticle current, Fig. S15d, do not affect the electric current, Fig. S15c, which is entirely determined by its Ohmic (harmonic) component.

We would also like to notice that the association of the sign of the electric charge density $\delta n$ with the sign of the variation of the quasiparticle density $\delta \rho$ and the direction of the electric $\boldsymbol{j}$ and quasiparticle $\boldsymbol{P}$ currents depends on the relative magnitude of the electric conductivities $\sigma_\alpha$ and the diffusive coefficients $D_\alpha$ of the electrons and holes, $\alpha = $ e, h. This property is demonstrated in the one-dimensional example shown in Fig. S12. Having the experimental access only to the charge density observable, we cannot determine the sign of the quasiparticle charge accumulation in Fig. S15b as well as the direction (positive or negative) of the associated quasiparticle currents in Fig. S15d.

## B. Dependence on lengths scales and boundary condition

Our simulations indicate that the size and the position of the pseudo-hydrodynamic vortices in the quasiparticle current depend not only on the recombination length $\lambda_R$ but also on the size of the electrodes, the distance to the boundaries and, to a smaller extent, on other parameters of the system such as the boundary parameter $g$. The same statement is also valid for the quasiparticle backflow, which determines the sign-alternating pattern of the experimentally observable charge accumulation near the electrodes. In order to make our discussion simpler and exclude one parameter, the electrode size, from our considerations, we discuss below the case of a pointlike electrode. The relevant formulae were already obtained in Section S12 C.

In Fig. S16 we show the charge density and the streamlines of the quasiparticle currents in the region very near to the pointlike electrode at the bottom edge of the sample. The behavior of these quantities, shown for a set of the inverse lengths $\varkappa_1$ and $\varkappa_2$, indicates that the size of the domains featuring the sign-alternating density and the size of the whirlpool in the quasiparticle current is set by the shortest length in the system. Additionally, we observe that at a fixed boundary condition ($g = 0$), the size of the whirlpool is correlated with the size of the induced charge density domain (the former is approximately twice smaller than the latter at $g = 0$). A similar quantitative picture is also observed for the quasiparticle density (not shown). The mechanism is the same: the backflow originating from the pseudo-hydrodynamic whirlpool brings the neutral quasiparticles back to the boundary and creates, due to the imbalance in the electron and hole mobilities, sign-alternating density patterns in electric charge that we observe in our optical experiment.

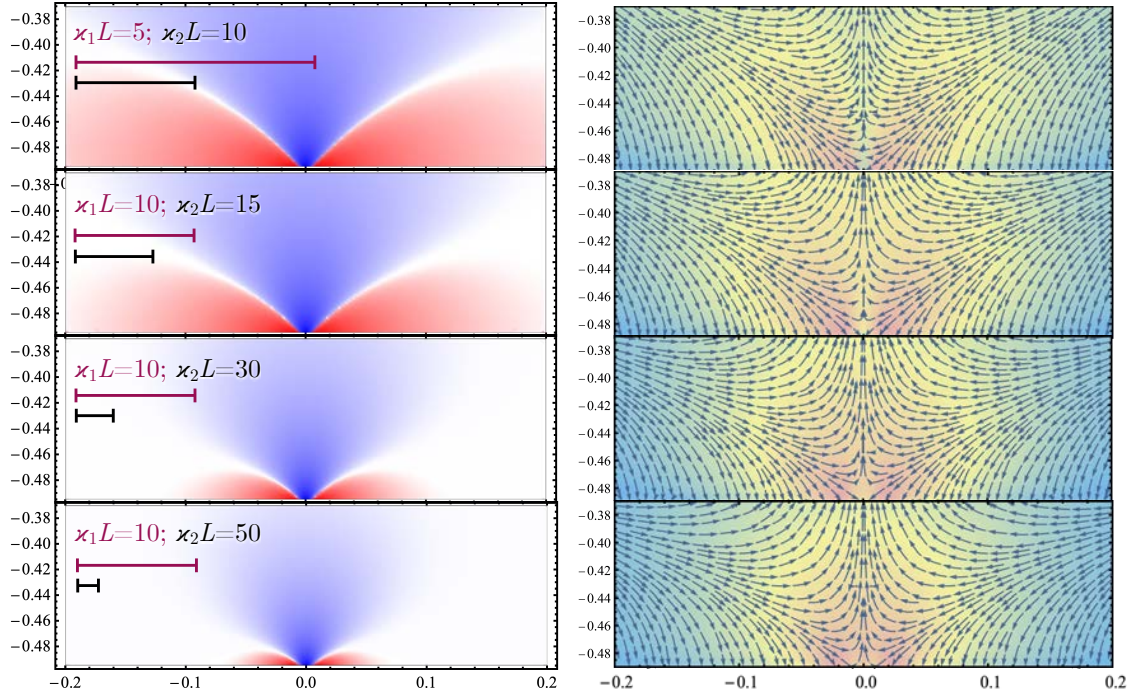

FIG. S16. The charge density (left) and the non-Ohmic component of the quasiparticle current (right) patterns close to the pointlike bottom electrode at various values of the inverse lengths $\varkappa_1$ and $\varkappa_2$ at the boundary condition $g = 0$ for the neutral current $\boldsymbol{P}$, Eq. (S132). The lengths $1/\varkappa_1$ and $1/\varkappa_2$ are shown graphically at the insets by the segments in the magenta and black colors.

The size of the charge-alternating domains depends on the longest correlation length in the system and the boundary condition. In Fig. S17 we show the same quantities as in Fig. S16 but for three different values of $g$ at fixed $\varkappa_1$ and $\varkappa_2$. While the size of the quasiparticle whirlpool stays unchanged, the size of the charge domain becomes substantially larger as the value of the coupling $g$ increases.

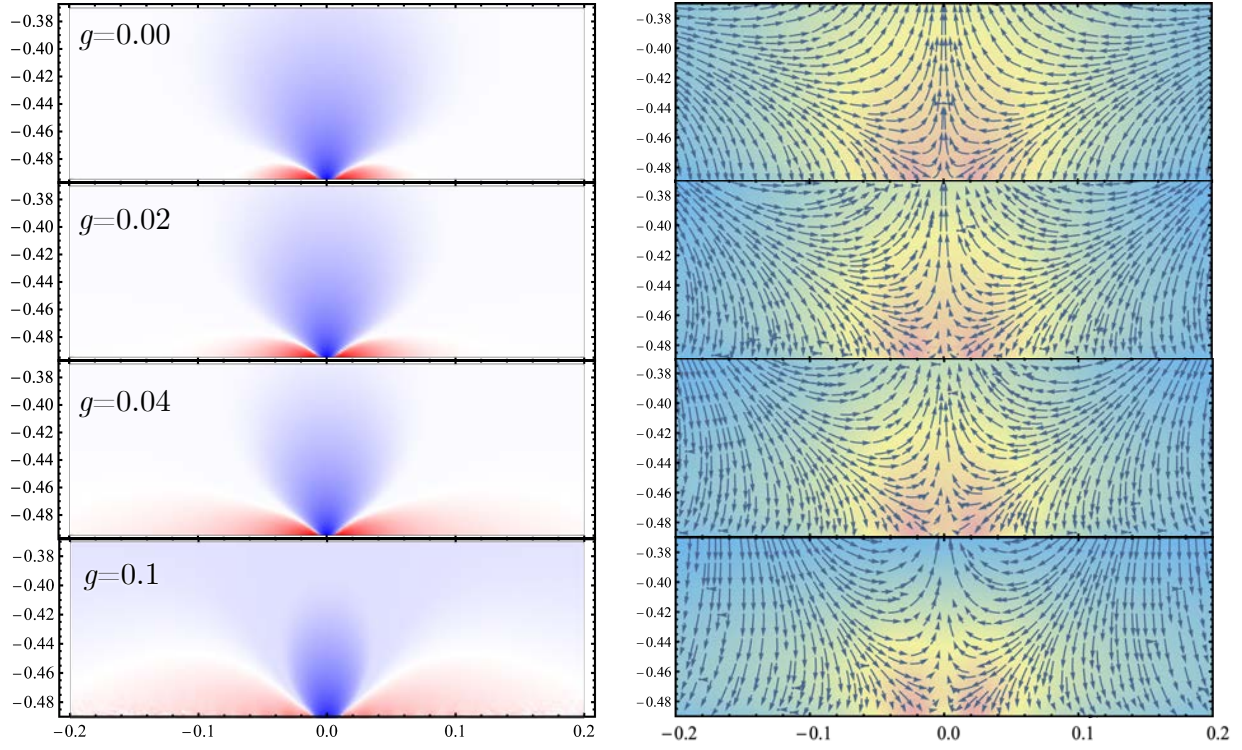

FIG. S17. The same as in Fig. S16 but for different constants $g = 0$, $0.02$, $0.04$, $0.1$ (from top to bottom) in the boundary conditions Eq. (S132) for the neutral current $\boldsymbol{P}$ at the fixed inverse lengths $\varkappa_1 L = 10$ and $\varkappa_2 L = 50$.

# Supplementary Materials References (Experiment)

# Supplementary Materials References (Theory)